\documentclass[twocolumn,english,aps,prb,showpacs,superscriptaddress]{revtex4-2}
\usepackage[T1]{fontenc}
\usepackage{amsmath,bm,graphicx,amssymb,epsfig,babel,dsfont}
\usepackage{mathrsfs}
\usepackage{color}

\newcommand{\rd}{{\rm d}}

\newcommand{\ri}{{\rm i}}

\newcommand{\re}{{\rm e}}

\begin{document}

\title{Goos-H\"{a}nchen effect singularities in transdimensional plasmonic films}

\author{Svend-Age Biehs}\email[Email: ]{s.age.biehs@uni-oldenburg.de}
\affiliation{Institut f\"{u}r Physik, Carl von Ossietzky Universit\"{a}t, 26111, Oldenburg, Germany}

\author{Igor V. Bondarev}\email[Email: ]{ibondarev@nccu.edu}
\affiliation{Department of Mathematics \& Physics, North Carolina Central University, Durham, NC 27707, USA}


\begin{abstract}
We identify and classify topologically protected singularities for the reflection coefficient of transdimensional plasmonic systems. Originating from nonlocal electromagnetic response due to vertical electron confinement in the system, such singularities lead to lateral (angular) Goos-H\"{a}nchen shifts on the millimeter (milliradian) scale in the visible range, greatly exceeding those reported previously for artificially designed metasurfaces, offering new opportunities for quantum material development.
\end{abstract}

\maketitle

%
%

It has been known that the reflection of a linearly polarized optical beam of finite transverse extent incident on a plane surface does not exactly follow the Snell's law of geometrical optics~\cite{Jackson99,BliokhAiello13}. Instead, the reflected beam experiences slight lateral in-plane displacement and angular deflection in the plane of incidence---the phenomenon commonly referred to as the Goos-H\"{a}nchen (GH) effect. Originating from the spatial dispersion of reflection or transmission coefficients due to the finite transverse size of the beam (and so nonlocal in its nature), the GH effect occurs for both reflected and refracted light in realistic optical systems. It was observed in a variety of systems (see Ref.\cite{BliokhAiello13}) including plasmonic metamaterials~\cite{GHmetamat}, graphene~\cite{GHgraphene} and even neutron scattering experiments~\cite{GHneutron}. These days the effect attracts much attention as well~\cite{GHbio24,Zubairy20,graphene18} due to the new generation of materials being available---quantum nanomaterials of reduced dimensionality, materials that can enhance nonlocal subwalength light propagation to offer new directions for quantum optics, quantum nanophotonics, and quantum computing application development~\cite{Sentef2022,GVidal2024}.

Plasmonic transdimensional (TD) quantum materials are atomically-thin metal (or semiconductor) films of precisely controlled thickness~\cite{BoltShalACS19}. Currently available due to great progress in nanofabrication techniques~\cite{Zayatz24,NL22TiN,GarciaAbajoACS19,thinXenes}, such materials offer high tailorability of their electronic and optical properties not only by altering their chemical and/or electronic composition (stoichiometry, doping) but also by merely varying their thickness (number of monolayers)~\cite{ManjavacasNatCom14,GarciaAbajoFD15,Brongersma17,BondOMEX17,BondMRSC18,BondOMEX19,TunableGold,BondPRR20,CNArr21PRAppl,Bond2022,Manjavacas22,BiehsBond2023,Shen2023,Bondarev2023,Pablo2024}. They provide a new regime---transdimensional, in between three (3D) and two (2D) dimensions, turning into 2D as the film thickness tends to zero. In this regime the strong vertical quantum confinement makes the linear electromagnetic (EM) response of the TD film nonlocal, or spatially dispersive, and the degree of nonlocality can be controlled by varying the film thickness~\cite{BondOMEX17,BondPRR20}. That is what makes plasmonic TD films indispensable for studies of the nonlocal light-matter interactions at the nanoscale~\cite{Bond2022,Manjavacas22,BiehsBond2023,Shen2023,Bondarev2023,Pablo2024}.

The properties of the TD plasmonic films can be explained by the confinement-induced nonlocal EM response theory~\cite{BondOMEX17,BondMRSC18} built on the Keldysh-Rytova (KR) electron interaction potential~\cite{KRK}. The theory is verified experimentally in a variety of settings~\cite{NL22TiN,Lavrinenko19,Shen2023}. It accounts for vertical electron confinement due to the presence of substrate and superstrate of dielectric permittivities less than that of the film, whereby the thickness of the film becomes a parameter to control its nonlocal EM response. The KR model covers both ultrathin films of thickness less than the half-wavelength of incoming light radiation and conventional films as thick as a few optical wavelengths~\cite{BondOMEX17,BondMRSC18}. The nonlocal EM response of TD plasmonic systems enables a variety of new effects such as thickness-controlled plasma frequency red shift~\cite{NL22TiN}, low-temperature plasma frequency dropoff~\cite{Lavrinenko19}, plasma mode degeneracy lifting~\cite{BondPRR20}, a series of quantum-optical~\cite{Bond2022,BondCN2024} and nonlocal magneto-optical effects~\cite{BondMRSC18} as well as thermal and vacuum field fluctuation effects responsible for near-field heat transfer~\cite{BiehsBond2023,Shen2023} and Casimir interaction phenomena~\cite{Bondarev2023,Pablo2024}.

Here, we focus on the confinement-induced nonlocality of the EM response of the TD plasmonic film to study theoretically the GH effect for an incident laser beam. Using the nonlocal KR model, analytical calculations and numerical analysis, we identify and classify topologically protected singularities for the nonlocal reflection coefficient of the system. Such singularities are shown to lead to giant lateral and angular GH shifts in the millimeter and milliradian range, respectively, to greatly exceed those of microscale reported for beams of finite transverse extent with no material-induced nonlocality~\cite{BliokhAiello13,GHmetamat,GHgraphene,GHneutron,GHbio24,Zubairy20}. They appear in TD materials with broken in-plane reflection symmetry (substrate and superstrate of different dielectric permittivities) where due to the confinement-induced nonlocality the eigenmode degeneracy is lifted to create the points of topological darkness in the visible range not existing in standard local Drude materials.

%
%

\begin{figure}[t]
\includegraphics[width=0.45\textwidth]{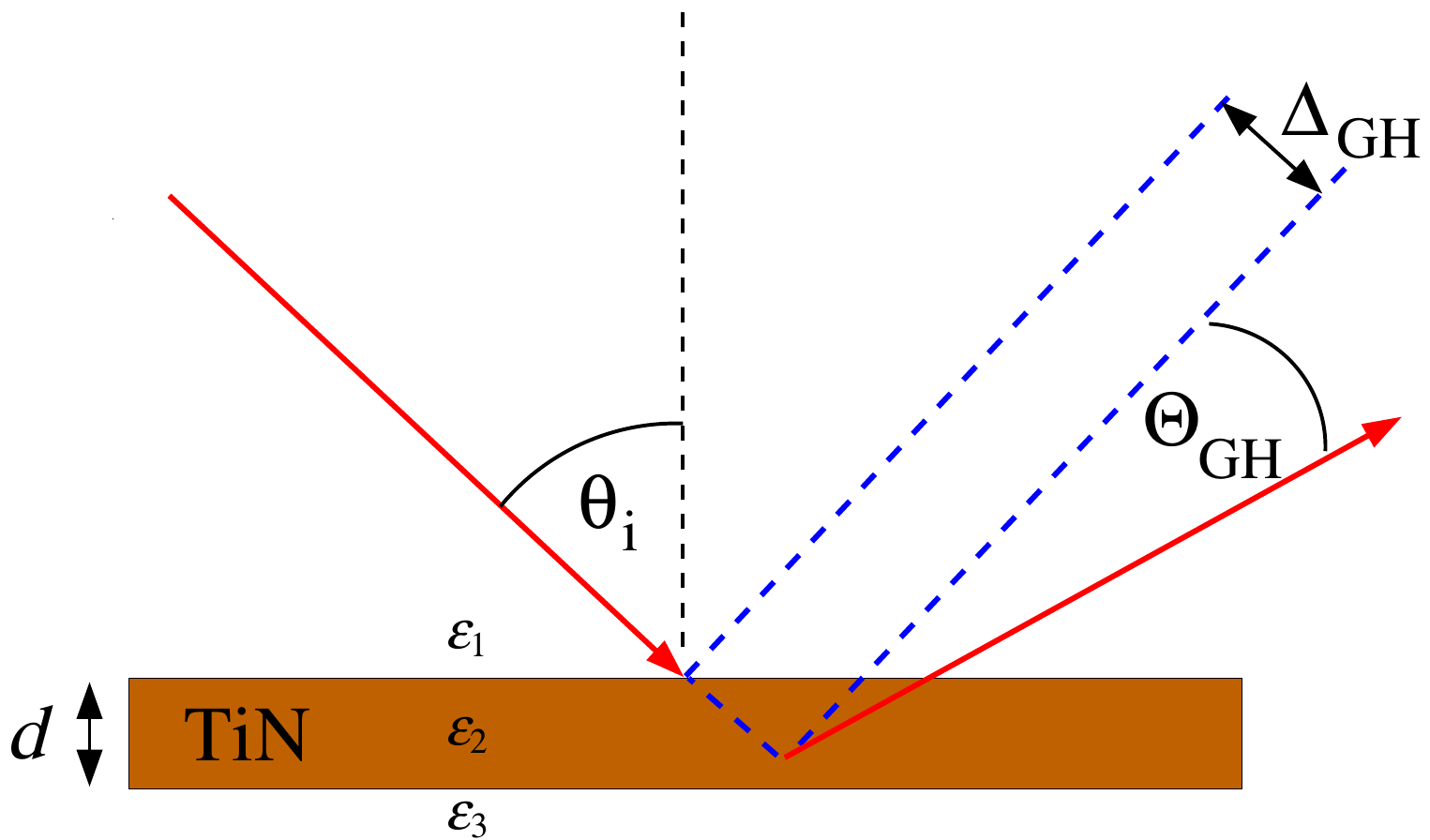}
\caption{\label{fig1}GH shifts $\Delta_{\rm GH}$ and $\Theta_{\rm GH}$ with TiN plasmonic slab. A detector placed a distance $l$ from the slab surface measures the shift $\Delta_{\rm total}=\Delta_{\rm GH} + l \tan(\Theta_{\rm GH}) \approx \Delta_{\rm GH} + l \Theta_{\rm GH}$.}
\end{figure}

The theory of the GH shift was originally formulated by Artmann back in 1948~\cite{Artmann}. The geometry of the effect is shown in Fig.~\ref{fig1}. After reflectance at an interface, the lateral and angular shifts of an incoming $p$-polarized wave in medium $1$ (refractive index $n_1$) are given by~\cite{Artmann,Aiello2009,BliokhAiello13,GH}
\begin{equation}
\Delta_{\rm GH} = n_1 \cos\theta_i\frac{\partial \varphi_p}{\partial k}
\label{Delta}
\end{equation}
and
\begin{equation}
\Theta_{\rm GH} = -\frac{\theta_0^2}{2} k_0 n_1 \frac{\cos\theta_i}{|R_p|}\frac{\partial |R_{\rm p}|}{\partial k},
\label{theta}
\end{equation}
respectively. Here, $k\!=\!k_0n_1\sin\theta_i$ is the wavevector in-plane projection, $\theta_i$ is the angle of incidence, $k_0\!=\!\omega/c$, and $\theta_0\!=\!2/(w_0k_0n_1)$ is the angular spread of an incident Gaussian light beam of waist $w_0$. The $p$-wave reflection coefficient is written as $R_p\!=\!|R_p|\re^{\ri\varphi_p}$ in the complex exponential form. Both shifts can be seen being spatially dispersive, with $\Delta_{\rm GH}$ being sensitive to reflectivity phase jumps and $\Theta_{\rm GH}$ to zero reflection itself so that large effects are highly likely for all kinds of zero reflection modes in the system. Phase jumps and singularities make the phase ill-defined, in which case the reflection coefficient absolute value must go to zero for causality reasons. In nonlocal materials such as our TD plasmonic films, phase jumps and singularities come from material spatial dispersion in addition to that of the light beam itself. This is precisely what we study here. The respective extra terms are derived for the structure in Fig.~\ref{fig1} and can be found elsewhere~\cite{SUPPL}. Expressions similar to Eqs.~(\ref{Delta}) and (\ref{theta}) can be written for $s$-polarized waves as well. However, as they show no peculiarities such as those we are about to discuss, we leave them out~\cite{SUPPL}. The derivations of relevance can also be found in Refs.~\cite{Aiello2009,BliokhAiello13}.

\begin{figure}[t]
\includegraphics[width=0.45\textwidth]{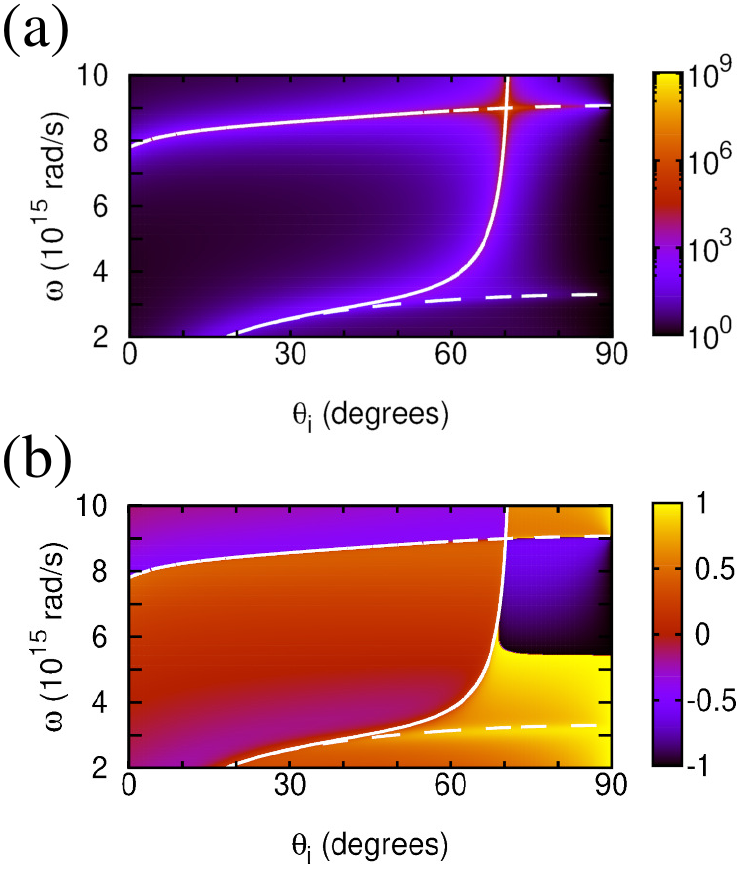}\vskip-0.25cm
\caption{\label{fig2}Inverse reflectivity $|R_{p}|^{-2}$ (a) and reflection phase $\varphi_p/\pi$ (b) for the $40\,{\rm nm}$ thick free-standing nonlocal TiN film. Shown are the $(n\!=\!1)$-SW (upper dashed line), the nonlocal CP (lower dashed line), and the BM (solid line).}
\end{figure}

%
%

For a free standing plasmonic film of thickness $d$ in air (Fig.~\ref{fig1}), the $p$-polarized wave reflection coefficient is~\cite{YEH}
\begin{equation}
R_p = \frac{r_p^{12} + r_p^{23} \re^{2 \ri \gamma_2 d}}{1 + r_p^{12} r_p^{23} \re^{2 \ri \gamma_2 d}},
\label{Rp}
\end{equation}
with medium 1 (superstrate) and medium 3 (substrate) having the same permittivities $\epsilon_1\!=\epsilon_3\!=\!1$. For medium~2 (film) we use $\epsilon_2\!=\!\epsilon_{\rm TiN}$ taking a TiN example of TD material that surpasses noble metals such as Au and Ag~\cite{ABoltSci}. The latter have exceptional plasmonic properties but relatively low melting temperatures making them incompatible with semiconductor fabrication technologies. On the contrary, transition metal nitrides have low-loss plasmonic response, high melting point and structural stability that makes them capable of forming stoichiometrically perfect TD films down to $1$~nm in thickness at room temperature~\cite{NL22TiN,SciAdv24}. The Fresnel reflection coefficients $r_p^{ij}$ ($i,j = 1,2,3$) for interfaces between medium 1 and 2 and between medium 2 and 3 are defined as follows
\begin{equation}
r_p^{ij} = \frac{\gamma_i \epsilon_j - \gamma_j \epsilon_i}{\gamma_i \epsilon_j + \gamma_j \epsilon_i},
\label{rp}
\end{equation}
where $\gamma_i = \sqrt{k_0^2 \epsilon_i - k^2}$ are the wave vectors components normal to the interface. Here, $r^{23}_p\!=\!-r_p^{12}$ as $\epsilon_1\!=\epsilon_3\!=\!1$, in which case zeroes of $R_p$ are determined by the Brewster mode (BM) condition $r^{12}_p\!=0$ at the film-air interface, whereby $\gamma_1 \epsilon_2 = \gamma_2 \epsilon_1$, leading to the dispersion relation
\begin{equation}
k = \frac{\omega}{c} \sqrt{\frac{\epsilon_{\rm TiN}'}{\epsilon_{\rm TiN}' + 1}}.
\label{Eq:Brewster}
\end{equation}
Also, zeros of $R_p$ can come from the film standing wave (SW) condition, $1 - \exp(2 \ri \gamma_2 d) = 0$, in which case
\begin{equation}
k = \sqrt{ \frac{\omega^2}{c^2} \epsilon_{\rm TiN}' - \biggl(\frac{n \pi}{d} \biggr)^2}.
\label{Eq:Standingwaves}
\end{equation}
Here, $n\!=\!1,2,3,...$ and $\epsilon_{\rm TiN}'\!=\!\mbox{Re}\,\epsilon_{\rm TiN}$. Lastly, zero reflection can also occur at the Christiansen point (CP) where $\epsilon_{\rm TiN}'\!=\!1$. If one uses the local Drude model (a 'workhorse' routinely used in plasmonics), then $\omega_{\rm CP}\!=\!4\!\times\!10^{15}\,{\rm rad/s}$ comes out of it for any angle of incidence, whereas in the nonlocal KR model used here the confinement-induced nonlocality of the EM response function $\epsilon_{\rm TiN}(\omega,k)$ makes the CP depend on $k$ and thus on the incidence angle. The comparison of the two as well as generally accepted TiN material parameters we use can be found elsewhere~\cite{SUPPL}.

\begin{figure}[t]
\includegraphics[width=0.45\textwidth]{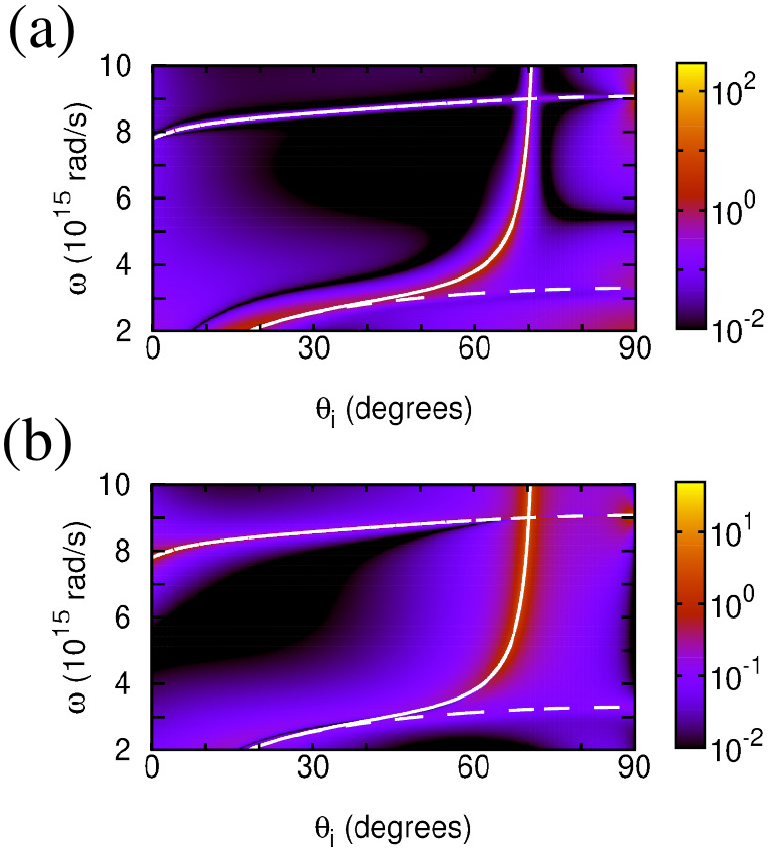}\vskip-0.25cm
\caption{\label{fig3}$|\Delta_{\rm GH}|$ in $\mu$m (a) and $|\Theta_{\rm GH}|$ in mrad (b) calculated for $p$-polarized lightwave incident on the 40~nm thick TiN film with inverse reflectivity and reflection phase shown in Fig.~\ref{fig2}.}
\end{figure}

Figure~\ref{fig2} shows the inverse reflectivity function $1/|R_p|^2$ and reflection phase $\varphi_p/\pi$ calculated from Eqs.~(\ref{Rp}) and (\ref{rp}) for an example of the 40~nm thick free standing TiN slab with nonlocal EM response. Reflectivity reduction and phase jumps can be seen when the SW, BM or CP mode are excited in the system. The BM still exists below the bulk TiN plasma frequency $\omega_p^{\rm 3D}\!=\!3.8\times10^{15}\,{\rm rad/s}$, making the GH shifts observable for $\omega\!<\!\omega_p^{\rm 3D}$, including the visible range that is impossible to access with local Drude-like plasmonic materials. Figure~\ref{fig3}, calculated from Eqs.~(\ref{Delta}) and (\ref{theta}) for the same example, shows strong lateral and angular GH shifts when the BM is excited, up to  $\approx80\,\mu{\rm m}$ and $\approx 30\,{\rm mrad}$, respectively. TD plasmonic materials thereby open access to the GH effect observation with visible light due to their remarkable property of the confinement-induced nonlocal EM response.

\begin{figure}[t]
\includegraphics[width=0.45\textwidth]{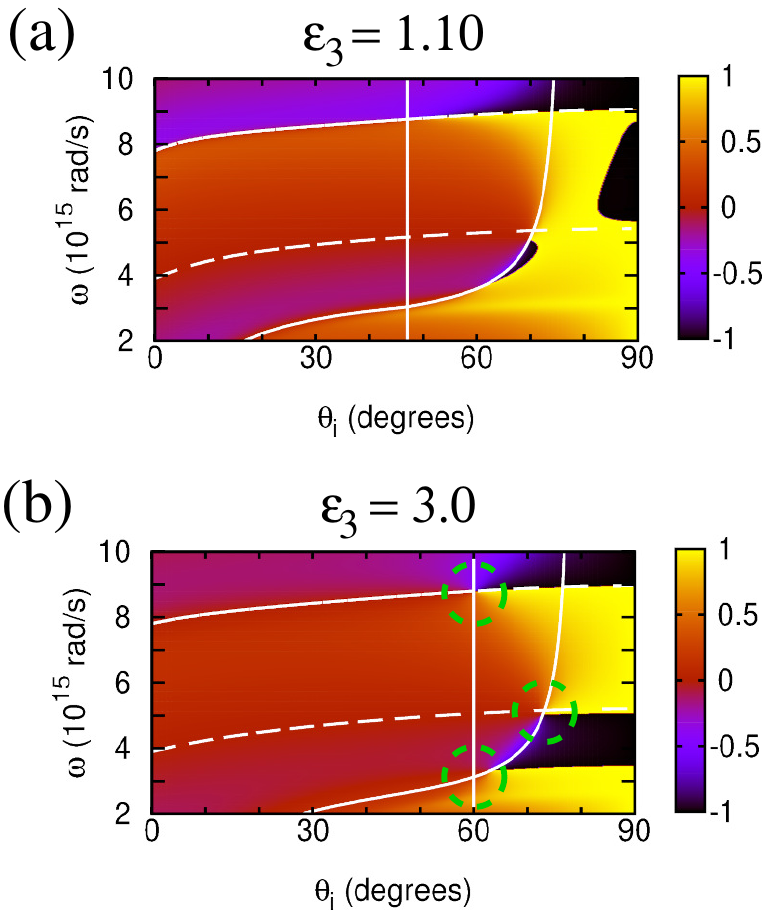}\vskip-0.25cm
\caption{\label{fig4}Reflection phase $\varphi_p/\pi$ calculated for a TiN film of $d = 40\,{\rm nm}$ with varied $\epsilon_3$ (substrate) and $\epsilon_1=1$ (superstrate). Vertical line is the $d\!=\!0$ BM, dashed lines are the first SWs of Eq.~(\ref{Eq:Standingwaves}) with $n\!=\!0.5,1$ and solid line is the gBM of Eq.~(\ref{Eq:GenBrewster}) (see text). Green circles mark the phase singularity points.}
\end{figure}

%
%

For TD plasmonic films sandwiched between superstrates and substrates with $\epsilon_1\!\ne\!\epsilon_3$, the in-plane reflection symmetry is broken and top-bottom interface mode degeneracy is lifted. Figure~\ref{fig4} shows the birth and development of phase singularities in this case as with $\epsilon_1\!=\!1$ (air) the substrate permittivity rises up to $\epsilon_3\!=\!3$ (MgO typically used in TiN thin film systems~\cite{NL22TiN}). It can be seen that just a slight substrate-superstrate dielectric permittivity difference gives birth to the two phase singularities close to the BM phase jump. They have opposite winding numbers (topological charges)
\begin{equation}
C = \frac{1}{2 \pi}\oint_\alpha\bm\nabla\varphi_p\cdot\rd\textbf{s}=\pm 1,
\end{equation}
where $\alpha$ is a closed integration path around the phase singularity point. There is also another phase singularity on the SW branch, which moves for larger substrate permittivities close to the SW and BM crossing point. Such singularities result in zero reflection previously reported as 'points of topological darkness' for specially designed metasurfaces~\cite{Grigorenko,Kravets} and ultrathin multilayer nanostructures~\cite{Singh, Ermolaev}. Here, we observe these topologically protected singularity points in mere nonlocal TD films. Remarkably, though, due to the plasma frequency decrease with thickness $d$ in our case~\cite{SUPPL,NL22TiN}, the gap in Fig.~\ref{fig4} between the low-frequency opposite topological charge singularities widens in thinner films (not shown), to red-shift the lower- and blue-shift the higher-frequency singularity points, respectively. By reducing $d$ in a controllable way in our case it can therefore be possible to access the low frequency phase singularity point with He-Ne laser at $\lambda = 632.8\,{\rm nm}$ ($3\times10^{15}\,{\rm rad/s}$) to observe an enhanced GH effect in the visible range. This remarkable feature is only offered by the TD plasmonic films due to their confinement-induced nonlocality and can never be realized with local Drude-like materials.

\begin{figure}[t]
\includegraphics[width=0.45\textwidth]{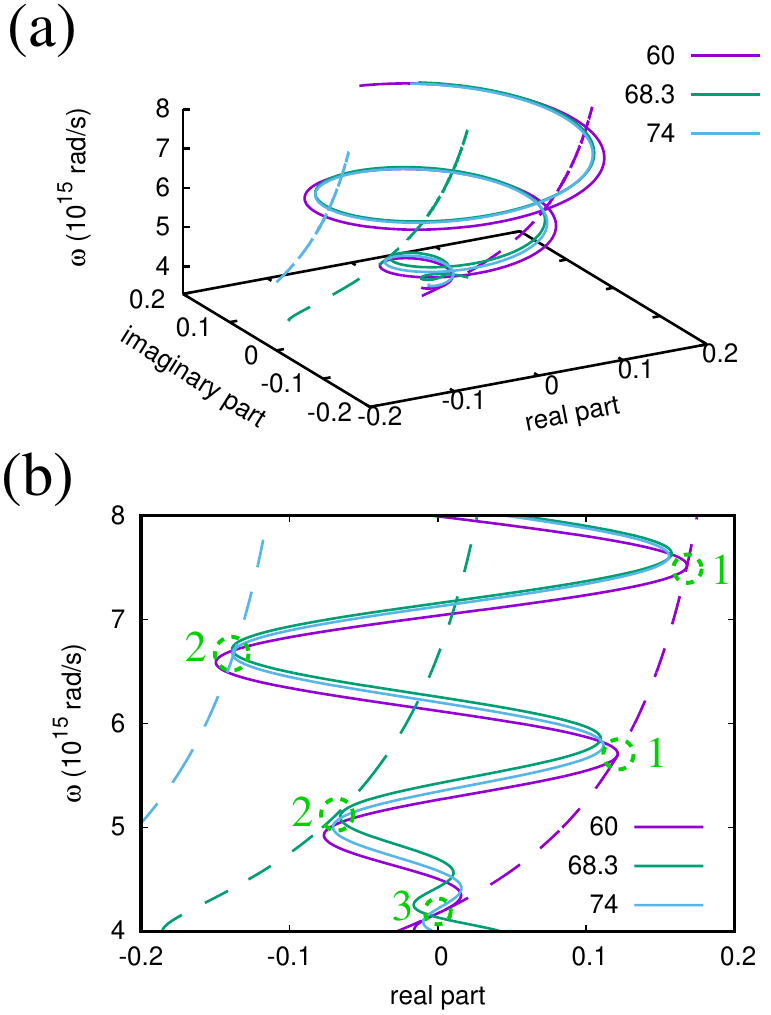}\vskip-0.25cm
\caption{\label{fig5}(a) Real and imaginary parts of the LHS (solid lines) and RHS (dashed lines) of Eq.~(\ref{Eq:circle}) for a few incidence angles. (b) Projection of (a) on the plane spanned by $\omega$ and real part. Angles are chosen to show the three solution cases to yield phase singularities in an air/TiN/MgO structure with $d\!=\!150\,{\rm nm}$. Green circles mark the phase singularity points.}
\end{figure}
\begin{figure}[t]
\includegraphics[width=0.5\textwidth]{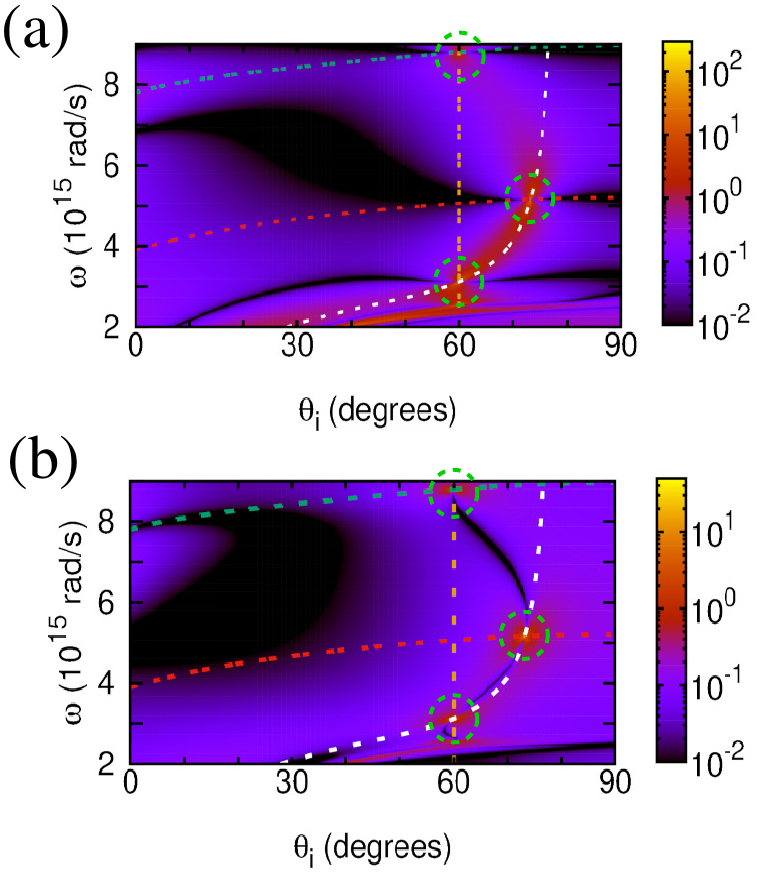}\vskip-0.25cm
\caption{\label{fig6}Absolute values of lateral (a) and angular (b) GH shifts $|\Delta_{\rm GH}|$ ($\mu$m) and $|\Theta_{\rm GH}|$ (mrad) calculated neglecting dissipation for the air/TiN/MgO system with $40$~nm thick TiN film. Green (red) dashed line is the SW (SQW) with $n\!=\!1$ ($n\!=\!0.5$) of Eq.~(\ref{Eq:Standingwaves}). Vertical orange dashed line is the zBM of Eq.~(\ref{Eq:Brewster13}). White dashed line is the gBM from Eq.~(\ref{Eq:GenBrewster}). Circles indicate the three cases of phase singularity points.}
\end{figure}

To understand the origin of the topological singularities discussed, it is instructive to represent the condition $R_p\!=\!0$ in the form
\begin{equation}
r_p^{23} \re^{2 \ri \gamma_2 d} = - r_p^{12}
\label{Eq:circle}
\end{equation}
with dissipation temporarily neglected. In this case the interface reflection coefficients are real numbers for all $\omega$ where $\epsilon_2(\omega,k)\!>\!0$ (and $\gamma_2\!>\!0$ accordingly) so that our TD film behaves as a dielectric. The LHS of this equation makes a circle in the complex plane of radius $r_p^{23}$ with phase $\theta\!=\!2\gamma_2 d$, while the RHS is a real number varying between $-1$ and $+1$ as $\omega$ and $\theta$ change. This is why the equality can only be achieved if the LHS is a real number as well, yielding two cases for Eq.~(\ref{Eq:circle}) to fulfill. They are: (1)~$\re^{2 \ri \gamma_2 d}\!=\!+1$, $r_p^{23}\!=\!-r_p^{12}$ and (2)~$\re^{2 \ri \gamma_2 d}\!=\!-1$, $r_p^{23}\!=\!r_p^{12}$. Here, the first equations are the SW condition $1\!-\re^{2\ri\gamma_2 d}\!=\!0$ and its alternative $1\!+\re^{2\ri\gamma_2 d}\!=\!0$. Both of them lead to the same dispersion relation of Eq.~(\ref{Eq:Standingwaves}), but the former for $n=1,2,3,\ldots$ and the latter for $n=0.5,1.5,2.5,\ldots$ describing SWs with multiples of a quarter wavelength and so to be referred to as standing quarter waves (SQW), accordingly. The second equation of case 1 is per Eq.~(\ref{rp}) fulfilled if
\begin{equation}
\gamma_1 \epsilon_3 = \epsilon_1 \gamma_3\,.
 \label{Eq:Brewster13}
\end{equation}
This is the substrate-superstrate interface BM condition, which can only be realized hypothetically for $d\!=\!0$, and so to be refereed to as zero-thickness BM (zBM). Similarly, the second equation of case (2) is fulfilled if
\begin{equation}
\gamma_2^2 \epsilon_1 \epsilon_3 = \epsilon_2^2 \gamma_3 \gamma_1.
 \label{Eq:GenBrewster}
\end{equation}
For $\epsilon_3\!=\!\epsilon_1$ this reduces to the symmetric in-plane interface condition leading to the BM dispersion relation of Eq.~(\ref{Eq:Brewster}). However, since it also holds for a broken in-plane reflection symmetry with $\epsilon_3\!\neq\!\epsilon_1$, we refer to the solution of this equation as the generalized BM (gBM).

The two simultaneous equations of cases 1 and 2 are represented by the lines in the 2D configuration space spanned by frequency and angle of incidence. The solutions to Eq.~(\ref{Eq:circle}) are given by the intersection of these lines. This is where the phase singularity points come from that we obtain at certain frequencies and incident angles. Additionally, there is a singularity coming from the trivial solution of Eq.~(\ref{Eq:circle}), to give case 3 where $r_p^{23}\!=r_p^{12}\!=0$, or $r_p^{23}\!=\!-r_p^{12}$ and $r_p^{23}\!=r_p^{12}$ simultaneously. This leads to the phase singularity at the intersection of the air-TiN and TiN-MgO interface BMs, or more generally, at the intersection of zBM and gBM lines in the configuration space. A complimentary analysis of the Christiansen points can be found elsewhere~\cite{SUPPL}.

Figure~\ref{fig5}~(a) shows the real and imaginary parts of Eq.~(\ref{Eq:circle}) calculated for a thicker TiN film of $d\!=\!150$~nm (to include more solution points) as functions of frequency for a few incident angles fixed. In Fig.~\ref{fig5}~(b) that presents a projection of (a), solution cases 1, 2 and 3 discussed above are marked accordingly. Here, case 1 can be seen to generate an infinite number of discrete solution points for a single 60$^\circ$ incident angle as $\epsilon_1$ and $\epsilon_3$ in Eq.~(\ref{Eq:Brewster13}) are frequency independent constants. Case 2 yields the solution points at different incident angles 68.3$^\circ$, 74$^\circ$, and 76$^\circ$ (not shown), etc. (also an infinite number, in principle) due to the strong gBM frequency dependence in Eq.~(\ref{Eq:GenBrewster}). The case 3 solution point can be seen at the intersection of the two lines of the 60$^\circ$ incidence angle, to yield the phase singularity point that can be shifted down to the visible $\omega$ for thinner films (see below) due to the confinement-induced nonlocal EM response.

\begin{figure}[t]
\includegraphics[width=0.45\textwidth]{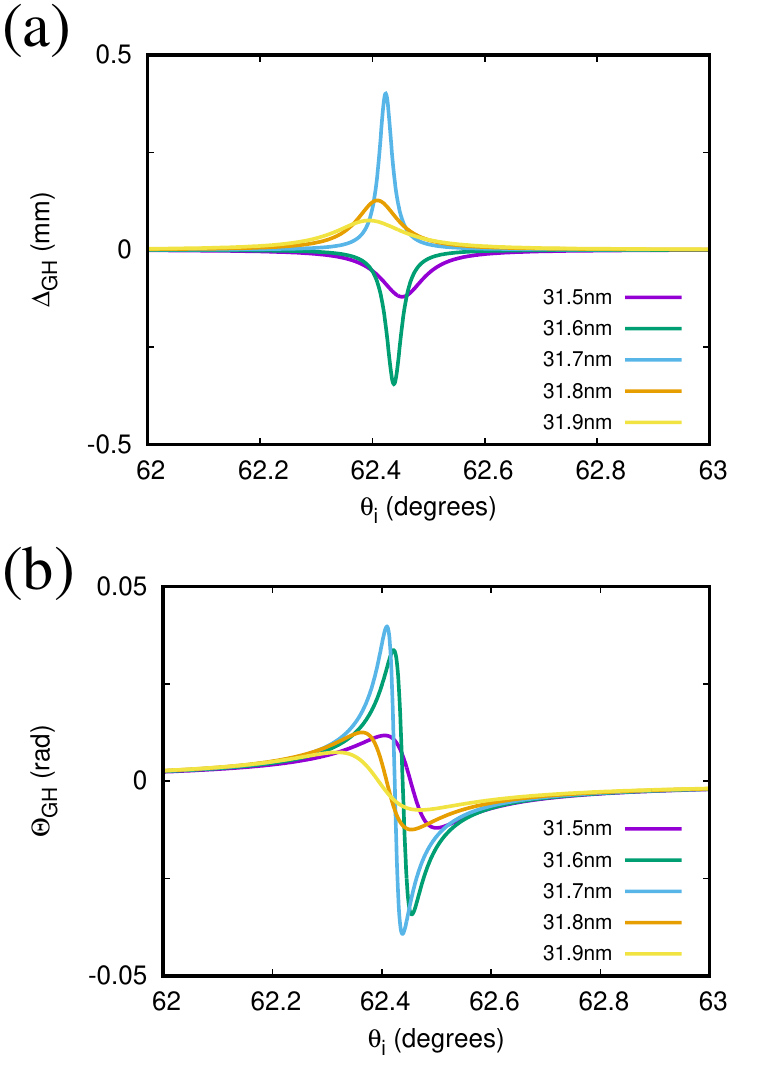}\vskip-0.5cm
\caption{\label{fig7}Lateral (a) and angular (b) GH shifts $\Delta_{\rm GH}$ (mm) and $\Theta_{\rm GH}$ (rad) calculated for the nonlocal dissipative ultrathin air/TiN/MgO system with TiN thickness in the vicinity of $d\!=\!31.7\,{\rm nm}$ and $\theta_i\!=\!62.42^\circ$ (case 3 phase singularity point).}
\end{figure}

The impact of phase singularities on the GH shifts is shown in Fig.~\ref{fig6} for the air/TiN/MgO system of the $40$~nm thick TiN film with dissipation neglected. Comparing to Fig.~\ref{fig4}~(b), the GH shifts can be seen to be very pronounced at the phase singularity points, which move with dissipation just slightly (not shown), for all three cases discussed. Using thinner TD films with broken in-plane symmetry it is even possible to make these GH shift singularities observable in the visible range under He-Ne laser excitation ($\omega\!=\!3\times\!10^{15}\,{\rm rad/s}$). Figure~\ref{fig7} shows the GH shifts under such excitation with angle of incidence varied around $\theta_i\!=\!62.42^\circ$ for the same air/TiN/MgO system (dissipation included~\cite{SUPPL}) with TiN film thickness varied around $d\!=\!31.7\,{\rm nm}$ to encounter the case 3 phase singularity point. For $d$ slightly thicker or thinner than that the singularity shows up slightly red- or blue-shifted, depending on its topological charge, making $\Delta_{\rm GH}$ of Eq.~(\ref{Delta}) change its sign as $d$ decreases. Similarly, the sign of $\Theta_{\rm GH}$ of Eq.~(\ref{theta}) changes as the incidence angle varies around $\theta_i\!=\!62.42^\circ$ of the case 3 phase singularity point. Most important though is that both lateral and angular GH shifts shown are extremely large, in the millimeter and a few tens of millirad range, respectively, over an order of magnitude greater than those shown in Fig.~\ref{fig3} for a similar in-plane symmetric TD film system.


To summarize, we show that the TD plasmonic films with broken in-plane reflection symmetry can feature an extraordinarily large GH effect in the visible range due to the topologically protected phase singularities of their reflection coefficient. This stems from the confinement-induced EM response nonlocality of the system, with plasma frequency decreasing as the film gets thinner~\cite{NL22TiN}. The remarkable opportunity to bring the GH effect to the visible range comes from the nonlocal plasma frequency which can be red-shifted not only by thickness but also by in-plane momentum reduction through the incident angle change. Previous studies reported an overall GH shift of $325\,\mu{\rm m}$ by coupling incident light to a surface plasmon polariton of a plasmonic film~\cite{GH}, as well as lateral and angular GH shifts as large as $70$ times $785\,{\rm nm}$ incident wavelength and $200\,\mu{\rm rad}$, respectively, for artificially designed metasurface structures~\cite{Yallapragada2016}. Our analysis reveals even greater lateral and angular GH shifts $\sim\!0.4$~mm (632 times the wavelength) and $\sim\!40$~mrad for visible He-Ne laser light with simple TD plasmonic films, indicating that such systems could provide a new flexible quantum material platform to offer new opportunities for quantum optics, quantum computing and biosensing application development.

\begin{acknowledgments}
This research was supported in part by grant NSF PHY-2309135 to the Kavli Institute for Theoretical Physics (KITP). I.V.B. gratefully acknowledges support from the U.S. Army Research Office under award No. W911NF2310206.
\end{acknowledgments}

\end{document}


\title{\texttt{\huge{Supplementary Information}}\\[2.5cm] Goos-H\"{a}nchen Effect Singularities in Transdimensional Plasmonic Films\\[0.5cm]~}

\author{S.-A. Biehs}
\affiliation{Institut f\"{u}r Physik, Carl von Ossietzky Universit\"{a}t, 26111, Oldenburg, Germany}

\author{I. V. Bondarev}
\affiliation{Department of Mathematics \& Physics, North Carolina Central University, Durham, NC 27707, USA}

\begin{abstract}
\vskip1cm Here we provide more information on the local and confinement-induced nonlocal contributions to the Goos-H\"{a}nchen (GH) effect, to include the analytical derivation of the local and nonlocal parts of the reflection coefficient to contribute to the GH shift, the nonlocal linear electromagnetic response (EM) expression of the Keldysh-Rytova model, the discussion of its significance for transdimensional material structures, and the numerical parameters used for our calculations within the framework of this model. Also, more discussion is provided for the local and nonlocal GH shifts in free standing TiN films and TiN films on a MgO substrate, including an in-depth analysis of the phase singularities in terms of the Brewster modes, standing waves, and Christiansen points.
\end{abstract}

\maketitle

\vskip1.5cm

\tableofcontents

\newpage

\subsection{GH Shifts for Gaussian Light Beams}

In this section, we give a brief sketch of the derivation of the GH shifts. As generally accepted in the literature~\cite{Aiello2008,Hecht}, we assume incoming light to impinge from medium $1$ (refractive index $n_1$) on the surface of medium~2 in the form of an incident p-polarized Gaussian beam with the electric field vector component as follows ($k_0=\omega/c$)
\begin{equation}
\mathbf{E}^{\rm inc}_p(x_i, y_i,z_i) \propto
\re^{\ri k_{0}n_1 z_i - k_{0}n_1\frac{x_i^2 + y_i^2}{2(L + \ri z_i)}} \biggl( \hat{\mathbf{x}}_i - \ri \hat{\mathbf{z}}_i \frac{x_i}{L + \ri z_i}\biggr).
\label{field}
\end{equation}
Here, $(x_i, y_i,z_i)$ are the coordinates in the Cartesian reference frame formed by the orthonormal basis vector set $(\hat{\mathbf{x}}_i,\hat{\mathbf{y}}_i,\hat{\mathbf{z}}_i)$ attached to the center of the beam cross-section such that $\hat{\mathbf{z}}_i$ sets up its propagation direction and $\hat{\mathbf{x}}_i$ lies in the plane of incidence pointing in the direction off the surface, $L = k_0 w_0^2/2$ with $w_0$ representing the beam waist. In full analogy, by implementing the boundary conditions, a similar expression can be written down for $\mathbf{E}^{\rm refl}_p$ in the $(\hat{\mathbf{x}}_r,\hat{\mathbf{y}}_r,\hat{\mathbf{z}}_r)$ reference frame attached to the center of the cross-section of the reflected beam. The total GH shift can then be calculated as the mean value
\begin{equation}
\langle x_r \rangle = \frac{\int\rd x_r \int \rd y_r\, I(x_r, y_r,z_r) x_r}{\int\rd x_r \int \rd y_r\, I(x_r, y_r,z_r)}
\end{equation}
of the centroid displacement for the reflected beam~\cite{Aiello2008}, where $I(x_r, y_r,z_r)\propto |\mathbf{E}^{\rm refl}_p(x_r, y_r,z_r)|^2$ is the reflected beam intensity in the far-field regime. This yields
\begin{equation}
\langle  x_r \rangle = \frac{1}{k_0} \Im \frac{\partial \ln R_\rp}{\partial \theta_i} - \frac{z_r}{k_0 L}  \Re \frac{\partial \ln R_\rp}{\partial \theta_i}
\label{xr}
\end{equation}
with $R_\rp$ representing the Fresnel reflection coefficient for p-polarized light, whereby for a detector placed a distance $z_r = l$ above the interface one obtains the total GH shift as a sum of the lateral $\Delta_{\rm GH}$ and angular $\Theta_{\rm GH}$ shifts of the form
\begin{equation}
	\Delta_{\rm total} = \Delta_{\rm GH} + l \tan(\Theta_{\rm GH}) \approx \Delta_{\rm GH} + l \Theta_{\rm GH},
\label{shift}
\end{equation}
\begin{equation}
\Delta_{\rm GH} = n_1\cos(\theta_i) \Im\biggl[\frac{1}{R_{\rm p}} \frac{\partial R_{\rm p}}{\partial k} \biggr],\;\;\;\;\;
\Theta_{\rm GH} = -\frac{\theta_0^2}{2} k_0n_1 \cos(\theta_i) \Re\biggl[\frac{1}{R_{\rm p}} \frac{\partial R_{\rm p}}{\partial k}\biggr].
\label{shifts}
\end{equation}
Here, $\theta_0 = \lambda/(\pi w_0) = 2/(w_0k_0n_1)$ and the partial derivatives over $\theta_i$ are replaced by those over $k = k_0 \sin(\theta_i)$. An in-depth analysis of the GH expressions can be found in Refs.~\cite{Aiello2009},\cite{Bliokh}.

\subsection{Local and Material-Induced Nonlocal GH Shifts}

For a material with a local ($k$-independent) EM response the derivatives over $k$ in Eq.~(\ref{shifts}) can only be nonzero due to the cross-section inhomogeneity of the incoming light beam as can be seen from Eqs.~(\ref{field})--(\ref{xr}) and the general structure of the p-wave reflection coefficient~\cite{YEH},
\begin{equation}
R_p = \frac{r_p^{12} + r_p^{23} \re^{2 \ri \gamma_2 d}}{1 + r_p^{12} r_p^{23} \re^{2 \ri \gamma_2 d}},\;\;\;\;\;
r_p^{ij} = \frac{\gamma_i \epsilon_j - \gamma_j \epsilon_i}{\gamma_i \epsilon_j + \gamma_j \epsilon_i},\;\;\;\;\;
\gamma_i = \sqrt{k_0^2 \epsilon_i - k^2}
\label{Rp}
\end{equation}
($i,j=1,2,3$), written for a typical case of a finite-thickness material slab of thickness $d$ with local EM response $\epsilon_2=\epsilon_2(\omega)$ sandwiched between semi-infinite superstrate and substrate dielectrics of constant permittivities $\epsilon_1$ and $\epsilon_3$, respectively. The s-wave reflection coefficient $R_s$ can be obtained by replacing $r_p^{ij}$ with $r_s^{ij}=(\gamma_i - \gamma_j)/(\gamma_i + \gamma_j)$ in the above equations. Thus, in Eq.~(\ref{shifts}) one has
\begin{equation}
\Delta_{\rm GH} = \Delta_{\rm GH}^{\rm loc} =
n_1 \cos(\theta_i) \Im\biggl[\frac{1}{R_{\rm p}} \biggl(\frac{\partial R_{\rm p}}{\partial k}\biggr)_{\rm loc} \biggr],\;\;\;\;\;
\Theta_{\rm GH} = \Theta_{\rm GH}^{\rm loc} =
-\frac{\theta_0^2}{2} k_0 n_1 \cos(\theta_i) \Re\biggl[\frac{1}{R_{\rm p}} \biggl(\frac{\partial R_{\rm p}}{\partial k}\biggr)_{\rm loc}\biggr].
\end{equation}
For a nonlocal material $\epsilon_2=\epsilon_2(\omega,k)$, and then there are extra contributions to add to the above, those proportional to $\partial\epsilon_2(\omega,k)/\partial k$, whereby Eq.~(\ref{shifts}) takes the form
\begin{equation}
\Delta_{\rm GH} = \Delta_{\rm GH}^{\rm loc} + \Delta_{\rm GH}^{\rm nloc}(k),\;\;\;\;\;
\Theta_{\rm GH} = \Theta_{\rm GH}^{\rm loc} + \Theta_{\rm GH}^{\rm nloc}(k)
\label{DT}
\end{equation}
with additional nonlocal terms
\begin{equation}
\Delta_{\rm GH}^{\rm nloc}(k) = n_1 \cos(\theta_i) \Im\biggl[\frac{1}{R_{\rm p}} \biggl(\frac{\partial R_{\rm p}}{\partial k}\biggr)_{\rm nloc} \biggr],\;\;\;\;\;
\Theta_{\rm GH}^{\rm nloc}(k) = -\frac{\theta_0^2}{2} k_0 n_1 \cos(\theta_i) \Re\biggl[\frac{1}{R_{\rm p}} \biggl(\frac{\partial R_{\rm p}}{\partial k}\biggr)_{\rm nloc}\biggr],
\end{equation}
which do not appear when the material-induced nonlocality is neglected.

\subsection{Derivative of the Nonlocal Reflection Coefficient}

In order to calculate the GH shifts in Eq.~(\ref{DT}), the partial derivative of the reflection coefficient $R_p$ over $k$ should be obtained first. With its definition given by Eq.~(\ref{Rp}), following is the full list of equations we used to calculate $\partial R_p/\partial k$ with both local and nonlocal terms included.

Using the 'prime'-sign $(^\prime)$ to abbreviate the first-order partial derivatives over $k$ and starting with $r_p^{ij}$, one has
\begin{equation}
\frac{\partial r_p^{ij}}{\partial k} = \biggl(\frac{\partial r_p^{ij}}{\partial k}\biggr)_{\rm loc} + \biggl(\frac{\partial r_p^{ij}}{\partial k}\biggr)_{\rm nloc}
\label{rij}
\end{equation}
with
\begin{equation}
\biggl(\frac{\partial r_p^{ij}}{\partial k}\biggr)_{\rm loc} = - \frac{A_{ij}}{B_{ij}^2} B_{ij,\rm loc}' + \frac{1}{B_{ij}} A_{ij, \rm loc}',\;\;\;\;\;
\biggl(\frac{\partial r_p^{ij}}{\partial k}\biggr)_{\rm nloc} =- \frac{A_{ij}}{B_{ij}^2} B_{ij,\rm nloc}' + \frac{1}{B_{ij}} A_{ij, \rm nloc}'\,,
\end{equation}
where
\begin{align}
r_p^{ij} &= \frac{A_{ij}}{B_{ij}},\;\;\;\;\;A_{ij} = \gamma_i \epsilon_j - \gamma_j \epsilon_i,\;\;\;\;\;B_{ij} = \gamma_i \epsilon_j + \gamma_j \epsilon_i,\label{AB}\\
A_{ij,\rm loc}' &= k \biggl( \frac{\epsilon_i}{\gamma_j} - \frac{\epsilon_j}{\gamma_i}\biggr),\;\;\;\;\;
A_{ij,\rm nloc}' = \epsilon_i' \biggl( \frac{\epsilon_j k_0^2}{2 \gamma_i} - \gamma_j\biggr) - \epsilon_j' \biggl( \frac{\epsilon_i k_0^2}{2 \gamma_j} - \gamma_i\biggr), \\
B_{ij,\rm loc}' &= -k \biggl( \frac{\epsilon_i}{\gamma_j} + \frac{\epsilon_j}{\gamma_i}\biggr),\;\;\;\;\;
B_{ij,\rm nloc}' = \epsilon_i' \biggl( \frac{\epsilon_j k_0^2}{2 \gamma_i} + \gamma_j\biggr) + \epsilon_j' \biggl( \frac{\epsilon_i k_0^2}{2 \gamma_j} + \gamma_i\biggr).
\end{align}

The nonlocal term in Eq.~(\ref{rij}) can be seen to be proportional to $\epsilon_i^\prime$ which in our configuration comes from the nonlocal EM response $\epsilon_{i=2}=\epsilon_2(\omega,k)$ of medium 2 situated in between dielectric media with constant $\epsilon_1$ (superstrate) and $\epsilon_3$ (substrate). In view of this, the $\partial R_p/\partial k$ derivative splits into the local and nonlocal contributions as follows
\begin{equation}
	\frac{\partial R_p}{\partial k} = \biggl(\frac{\partial R_p}{\partial k}\biggr)_{\rm loc} + \biggl(\frac{\partial R_p}{\partial k}\biggr)_{\rm nloc}
\end{equation}
with
\begin{equation}
\biggl(\frac{\partial R_p}{\partial k}\biggr)_{\rm loc} = - \frac{C}{D^2} D_{\rm loc}' + \frac{1}{D} C_{\rm loc}',\;\;\;\;\;
\biggl(\frac{\partial R_p^{ij}}{\partial k}\biggr)_{\rm nloc} =- \frac{C}{D^2} D_{\rm nloc}' + \frac{1}{D} C_{\rm nloc}',
\end{equation}
where
\begin{equation}
R_p = \frac{C}{D},\;\;\;\;\;C = r_p^{12} + r_p^{23} \re^{2 \ri \gamma_2 d},\;\;\;\;\;D = 1 + r_p^{12} r_p^{23} \re^{2 \ri \gamma_2 d}
\end{equation}
and
\begin{align}
C_{\rm loc}' &= (r_p^{12})_{\rm loc}' + (r_p^{23})_{\rm loc}' \re^{2 \ri \gamma_2 d} -  r_p^{23} \re^{2 \ri \gamma_2 d} \frac{2 \ri k d}{\gamma_2},\;\;\;\;\;
C_{\rm nloc}' = (r_p^{12})_{\rm nloc}' + (r_p^{23})_{\rm nloc}' \re^{2 \ri \gamma_2 d} +  r_p^{23} \re^{2 \ri \gamma_2 d} \frac{\ri k_0^2 d}{\gamma_2} \epsilon_2',\\
D_{\rm loc}' &= \biggl[(r_p^{12})_{\rm loc}' r_p^{23} +  r_p^{12} (r_p^{23})_{\rm loc}' -  r_p^{12} r_p^{23} \frac{2 \ri k}{\gamma_2} \biggr]\re^{2 \ri \gamma_2 d},\;\;\;
D_{\rm nloc}' = \biggl[(r_p^{12})_{\rm nloc}' r_p^{23} +  r_p^{12} (r_p^{23})_{\rm nloc}' -  r_p^{12} r_p^{23} \frac{\ri k_0^2}{\gamma_2} \epsilon_2' \biggr]\re^{2 \ri \gamma_2 d}.
\end{align}
Here, it can be seen that due to the presence of $\gamma_2=\sqrt{k_0^2\epsilon_2 - k^2}$, as long as medium 2 is nonlocal the local contribution is formally nonlocal as well, being also contributed by the nonlocal term proportional to $\epsilon_2^\prime$. The latter does not exists if medium 2 is local, in which case the former is the only nonzero local contribution, thus justifying its name.

Equations similar to the above can also be obtained for the s-polarization. Redefining $A_{ij}$ and $B_{ij}$ of Eq.~(\ref{AB}) as
\begin{equation}
\bar{A}_{ij} = \gamma_i - \gamma_j,\;\;\;\;\;\bar{B}_{ij} = \gamma_i + \gamma_j,\;\;\;\;\;r_s^{ij} = \frac{\bar{A}_{ij}}{\bar{B}_{ij}},
\end{equation}
one obtains
\begin{equation}
\frac{\partial r_s^{ij}}{\partial k} = \biggl(\frac{\partial r_s^{ij}}{\partial k}\biggr)_{\rm loc} + \biggl(\frac{\partial r_s^{ij}}{\partial k}\biggr)_{\rm nloc}
\end{equation}
with
\begin{equation}
\biggl(\frac{\partial r_s^{ij}}{\partial k}\biggr)_{\rm loc} = - \frac{\bar{A}_{ij}}{\bar{B}_{ij}^2} \bar{B}_{ij,\rm loc}' + \frac{1}{\bar{B}_{ij}} \bar{A}_{ij, \rm loc}',\;\;\;\;\;
\biggl(\frac{\partial r_s^{ij}}{\partial k}\biggr)_{\rm s,nloc} =- \frac{\bar{A}_{ij}}{\bar{B}_{ij}^2} \bar{B}_{ij,\rm nloc}' + \frac{1}{\bar{B}_{ij}} \bar{A}_{ij, \rm nloc}'\,,
\end{equation}
where
\begin{align}
\bar{A}_{ij,\rm loc}' &= k \biggl( \frac{1}{\gamma_j} - \frac{1}{\gamma_i}\biggr),\;\;\;\;\;\;\;
\bar{A}_{ij,\rm nloc}' = \epsilon_i'\frac{k_0^2}{2 \gamma_i} - \epsilon_j' \frac{k_0^2}{2 \gamma_j}, \\
\bar{B}_{ij,\rm loc}' &= -k \biggl( \frac{1}{\gamma_j} + \frac{1}{\gamma_i}\biggr),\;\;\;\;\;\;\;
\bar{B}_{ij,\rm nloc}' = \epsilon_i' \frac{k_0^2}{2 \gamma_i} + \epsilon_j' \frac{ k_0^2}{2 \gamma_j}.
\end{align}
This gives
\begin{equation}
\frac{\partial R_s}{\partial k} = \biggl(\frac{\partial R_s}{\partial k}\biggr)_{\rm loc} + \biggl(\frac{\partial R_s}{\partial k}\biggr)_{\rm nloc}
\end{equation}
with
\begin{equation}
\biggl(\frac{\partial R_s}{\partial k}\biggr)_{\rm loc} = - \frac{\bar{C}}{\bar{D}^2} \bar{D}_{\rm loc}' + \frac{1}{\bar{D}} \bar{C}_{\rm loc}',\;\;\;\;\;
\biggl(\frac{\partial R_s^{ij}}{\partial k}\biggr)_{\rm nloc} =- \frac{\bar{C}}{\bar{D}^2} \bar{D}_{\rm nloc}' + \frac{1}{\bar{D}} \bar{C}_{\rm nloc}'\,,
\end{equation}
where
\begin{equation}
\bar{C} = r_s^{12} + r_s^{23} \re^{2 \ri \gamma_2 d},\;\;\;\;\;\bar{D} = 1 + r_s^{12} r_s^{23} \re^{2 \ri \gamma_2 d}
\end{equation}
and
\begin{align}
\bar{C}_{\rm loc}' &= (r_s^{12})_{\rm loc}' + (r_s^{23})_{\rm loc}' \re^{2 \ri \gamma_2 d} -  r_s^{23} \re^{2 \ri \gamma_2 d} \frac{2 \ri k d}{\gamma_2},\;\;\;\;\;
\bar{C}_{\rm nloc}' = (r_p^{12})_{\rm nloc}' + (r_s^{23})_{\rm nloc}' \re^{2 \ri \gamma_2 d} +  r_s^{23} \re^{2 \ri \gamma_2 d} \frac{\ri k_0^2 d}{\gamma_2} \epsilon_2',\\
\bar{D}_{\rm loc}'\!&=\!\biggl[(r_s^{12})_{\rm loc}' r_s^{23} +  r_s^{12} (r_s^{23})_{\rm loc}'\!-  r_s^{12} r_s^{23} \frac{2 \ri k}{\gamma_2} \biggr]\re^{2 \ri \gamma_2 d},\;\;\;
\bar{D}_{\rm nloc}'\!=\!\biggl[(r_s^{12})_{\rm nloc}' r_s^{23} +  r_s^{12} (r_s^{23})_{\rm nloc}'\!-  r_s^{12} r_s^{23} \frac{\ri k_0^2}{\gamma_2} \epsilon_2' \biggr]\re^{2 \ri \gamma_2 d}.
\end{align}

The set of equations above is used to numerically evaluate the GH shifts.

\subsection{Confinement-Induced Nonlocal EM Response and Experimental Parameters Used}

The electrostatic Coulomb field produced by confined remote charge carriers outside of their confinement region starts playing a perceptible role with the confinement size reduction~\cite{KRK,Louie09}. The Coulomb interaction of the charges confined is typically stronger than that in a homogeneous medium with the same dielectric permittivity due to the increased field contribution from outside dielectric environment with lower dielectric constant. That is why to describe the optical properties of TD plasmonic films we use the confinement-induced nonlocal EM response theory built on the Keldysh-Rytova (KR) electron interaction potential~\cite{BondOMEX17,BondMRSC18}. This theory applies perfectly to our configuration of an optically dense ultrathin metallic material slab in region 2 of thickness $d$ surrounded by semi-infinite dielectrics of constant permittivities $\epsilon_1$ (top) and $\epsilon_3$ (bottom), to result in the in-plane EM response of medium 2 (confined region)
\begin{equation}
\epsilon_2(\omega,k)=\epsilon_b \biggl[1 - \frac{\omega_p^2(k)}{\omega(\omega + \ri \Gamma_D)} \biggr].
\label{eps2}
\end{equation}
Here, $\epsilon_b\,(\gg\!\epsilon_1,\epsilon_3)$ is the constant background permittivity, $\Gamma_D$ is the damping constant, and the plasma frequency
\begin{equation}
\omega_p(k) = \frac{\omega_p^{\rm 3D}}{\sqrt{1 + 1/(\tilde{\epsilon} k d)}},\;\;\;\;\;\tilde{\epsilon} = \frac{\epsilon_b}{\epsilon_1 + \epsilon_3}
\label{omegapk}
\end{equation}
is nonlocal due to the vertical electron confinement, turning into $\omega_p^{\rm 3D}$ in the limit $d\rightarrow\infty$ to represent the constant bulk material plasma frequency screened of the standard local (Drude) EM response of 3D metals~\cite{BondOMEX17}.

This theoretical model is verified in a variety of experiments~\cite{SciAdv24,Shen2023,NL22TiN,Lavrinenko19}, which is why we choose to set up
\begin{equation}
\epsilon_2(\omega,k)=\epsilon_{\rm TiN}(\omega,k)
\label{epstin}
\end{equation}
in our studies with TiN material parameters measured experimentally and tabulated for convenience herein~\cite{NL22TiN}. The Table presented also includes other numerical parameters used. Note that the damping constant $\Gamma_D$ depends on the film thickness, in general; however, for the range of film thicknesses we used it coincides with the bulk value.

With Eqs.~(\ref{eps2})-(\ref{epstin}) the derivative $\epsilon_2'$ in the nonlocal reflection coefficient expressions above reads as follows
\begin{equation}
\epsilon_2'=\frac{\partial\epsilon_{\rm TiN}(\omega,k)}{\partial k}=-\frac{\epsilon_b\omega_p^2(k)}{k(\tilde{\epsilon}kd+1)\omega(\omega+\ri\Gamma_D)} =
\frac{\tilde{\epsilon} d \big[\epsilon_{\rm TiN}(\omega,k)-\epsilon_b\big]}{(\tilde{\epsilon} k d + 1)^2}.
\end{equation}
It can be seen not only being nonzero at finite $d$ but also being both positive and negative depending on the frequency of light, and it disappears when $d$ goes to infinity---as it should to make the EM response of thick films local in accord with the standard Drude model.

\begin{table}[t]
\begin{tabular}{|c|c|c|c|c|c|c|c|}
\hline
~~~$\epsilon_b$ (TiN)~~~ & ~~~$\epsilon_1$ (air) ~~~ & ~~~$\epsilon_3$ (MgO) ~~~ & ~~~$\omega_p^{\rm 3D}$ (TiN), eV ~~~ & ~~~$\Gamma_{\rm D}$~(TiN), eV ~~~ & ~~~$w_0$ (beam waist), $\mu$m ~~~\\
\hline 9.1 & 1.0 & 3.0 & 2.5 & 0.2 & 32 \\ \hline
\end{tabular}
\label{Tab:parameters}
\end{table}

\subsection{Preserved In-Plane Reflection Symmetry: TiN Film Free Standing in Air}

In this section we analyze the EM modes of a finite-thickness TiN film free standing in air, in order to be able to identify the modes contributing to the GH shifts. In what follows, the 'prime'-sign $(^\prime)$ and 'double prime'-sign $(^{\prime\prime})$ abbreviate the real $\mbox{Re}(...)$ part and imaginary $\mbox{Im}(...)$ part, respectively.

We start with the interface EM modes between two infinitely extended media~\cite{Greffet}. They are the Brewster mode and the surface mode. Both of them can be present at the two air/TiN interfaces separating media 1 and 3 (air above and below the TiN film) from medium 2 (the film itself), and both are described by the condition
\begin{equation}
	k = \frac{\omega}{c} \sqrt{\frac{\epsilon_{\rm TiN}}{\epsilon_{\rm TiN}+1}}.
	\label{Eq:Brewster}	
\end{equation}
This can be obtained from either $r_p^{12}=-r_p^{23}=0$ with $\epsilon_{\rm TiN}^\prime\geq0$ ($\omega\geq\omega_p$) or $r_p^{12}=-r_p^{23}=\infty$ with $\epsilon_{\rm TiN}^\prime<0$ ($\omega<\omega_p$) in Eq.~(\ref{Rp}), to give for media 1 and 3 (air) the Brewster mode in the propagating wave region $k\leq\omega/c$ and the surface mode in the evanescent wave region $k > \omega/c$, respectively. Both of them lead to $R_p=0$ according to Eq.~(\ref{Rp}) and thus to the stronger GH effect as per Eq.~(\ref{shifts}). To obtain the corresponding dispersion relations, Eq.~(\ref{Eq:Brewster}) should be solved with dissipation included for either the complex valued $k$ at a given real frequency $\omega$ or the complex valued $\omega$ at a given real valued in-plane momentum $k$. Depending on the physical situation either approach has to be used~\cite{Greffet}.

Neglecting the dissipation in Eq.~(\ref{Eq:Brewster}) leads to the idealized solution with real valued $k$ and $\omega$, which for nonlocal $\epsilon_{\rm TiN}(\omega,k)$ of Eqs.~(\ref{eps2})-(\ref{epstin}) with $\Gamma_D=0$ takes the following form
\begin{equation}
	\omega^2 = \frac{1}{2}\biggl[k^2 c^2 \frac{\epsilon_b + 1}{\epsilon_b} + \omega_p^2(k) \biggr] \pm \sqrt{\frac{1}{4}\biggl[k^2 c^2 \frac{\epsilon_b + 1}{\epsilon_b} + \omega_p^2(k) \biggr]^2\!-k^2 c^2 \omega_p^2(k)}\,.
	\label{Eq:Brewsternonlocal}
\end{equation}
Here, the $+(-)$ sign solution describes the Brewster (surface) mode in the propagating (evanescent) wave region. With $\omega_p^2(k)$ of Eq.~(\ref{omegapk}), in the limit $d\rightarrow\infty$ the well known local dispersion relations can be recovered~\cite{Greffet}.

In the finite-thickness films, in addition to the interface modes there are also standing waves to represent the eigen modes of the film itself. The standing wave modes are responsible for enhanced transmission and so reduced reflection of the film. For our nonlocal free standing TD plasmonic film of thickness $d$, the standard standing wave condition is $\gamma_{2} = \pi n / d$ with $n \in \mathds{N}$, the same as that studied for polaritonic waves in thin dielectric films previously~\cite{KliewerFuchs}, yielding
\begin{equation}
k = \sqrt{ \frac{\omega^2}{c^2} \epsilon_{\rm TiN} - \biggl(\frac{\pi n}{d} \biggr)^2},
\label{Eq:Standingwaves}	
\end{equation}
to give
\begin{equation}
\omega = \sqrt{\frac{k^2 c^2}{\epsilon_b} + \omega_p^2(k) + \frac{c^2}{\epsilon_b}\biggl( \frac{\pi n}{d} \biggr)^2}
\label{Eq:Standingwavesnonlocal}	
\end{equation}
for $\epsilon_{\rm TiN}(\omega,k)$ of Eqs.~(\ref{eps2})-(\ref{epstin}) with dissipation neglected as before. Again, in the limit $d\rightarrow\infty$ one obtains the local version of the standing wave solutions in the TD plasmonic film. These solutions are similar but not exactly the same as those reported previously for the local standing polaritonic waves~\cite{KliewerFuchs} due to the different in-plane EM responses of the metallic and dielectric thin films.

\begin{figure}
\includegraphics[width=0.75\textwidth]{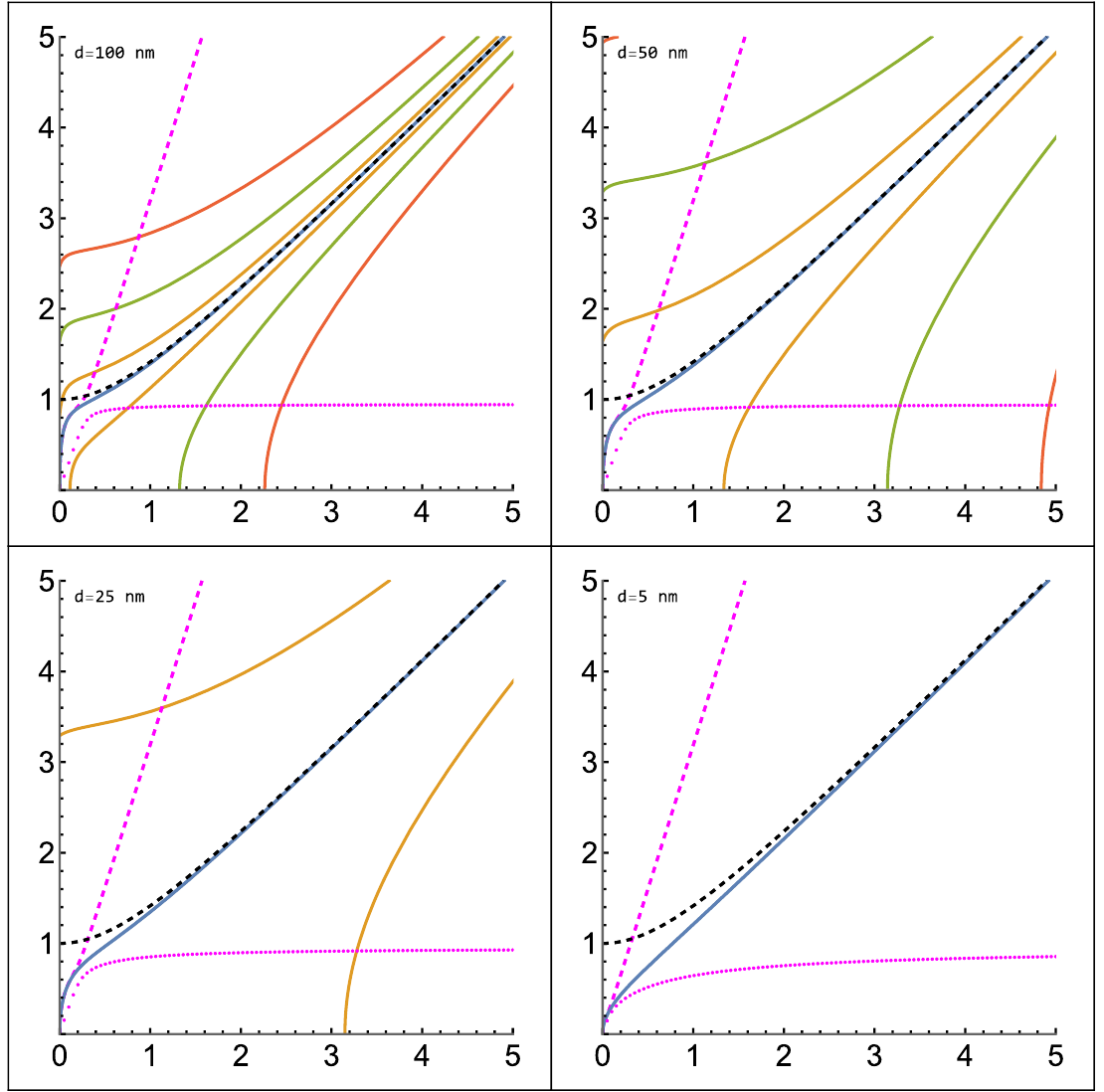}
\caption{Propagating and evanescent standing wave solutions with $n=0,1,2$ and $3$ calculated for the free standing TiN films of decreasing thickness $d$ from Eq.~(\ref{genstandingfin}) with frequency $\omega$ in units of $\omega_p^{\rm 3D}$ and in-plane momentum $k$ in units of $\sqrt{\epsilon_b}\,\omega_p^{\rm 3D\!}/c$. In a relatively thick film ($d=100$~nm), the $n\ne0$-modes can be seen to group around the fundamental plasma mode ($n=0$, shown by blue and black dashed line for nonlocal KR and local Drude in-plane EM response given by Eqs.~(\ref{eps2}) and (\ref{omegapk}) for finite and infinitely large $d$, respectively); propagating modes are above and evanescent modes are below the $n=0$-mode. In the limit $d\rightarrow\infty$, all of them congregate together to form the multiply degenerate bulk plasma mode with standard low-$k$ dispersion $\omega\!\sim\!\omega_p^{\rm 3D}$. As $d$ decreases, all $n\ne0$-modes spread out to reflect the strengthening of the vertical confinement in the TD film system. In ultrathin films ($d=5$~nm), only the fundamental $n=0$-mode remains with low-$k$ dispersion distinctly different for nonlocal and local in-plane EM response of the film ($\omega\!\sim\!\sqrt{k}$ versus $\omega\!\sim\!\omega_p^{\rm 3D}$)~\cite{BondOMEX17,BondPRR20}. Purple dashed and dotted lines indicate the Brewster and surface modes, respectively, of Eq.~(\ref{Eq:Brewsternonlocal}).}
\label{fig01}
\end{figure}

\begin{figure}
\includegraphics[width=0.49\textwidth]{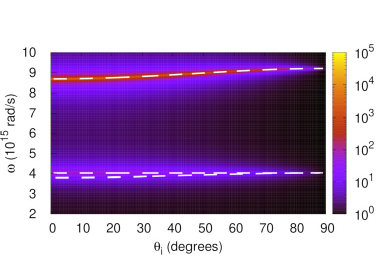}
\includegraphics[width=0.49\textwidth]{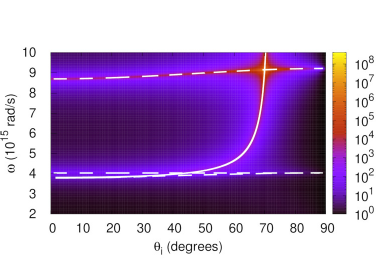}
\caption{\label{Fig:inverselocal}Inverse reflectivity $|R_{s/p}|^{-2}$ for s-polarized (left panel) and p-polarized (right panel) light calculated for a 40~nm thick TiN film using the local (Drude) in-plane EM response function one can obtain from Eqs.~(\ref{eps2}) and (\ref{omegapk}) in the limit $d\rightarrow\infty$. The white solid line is the Brewster mode of Eq.~(\ref{Eq:Brewsternonlocal}); it can be seen to only exist for p-polarized light (right panel). The horizontal dashed line at $\omega = 4\times10^{15}\,{\rm rad/s}$ indicates the Christiansen point (see the text below). Two other dashed lines shown are tilted, depend on the angle of incidence. They are the $n = 0$ fundamental plasma mode (lower dashed line) and the $n = 1$ standing wave mode (upper dashed line), both given by Eq.~(\ref{Eq:Standingwavesnonlocal}). For all modes mentioned the reflectivity can be seen to be close to zero for both s- and p-polarized light.}
\end{figure}

\begin{figure}
\includegraphics[width=0.49\textwidth]{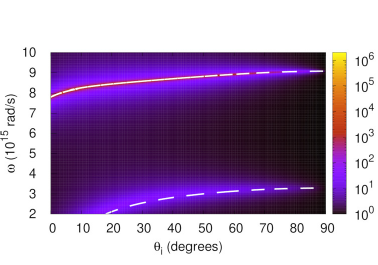}
\includegraphics[width=0.49\textwidth]{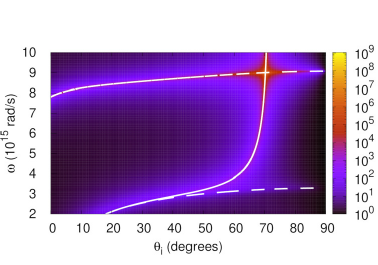}
\caption{\label{Fig:inversenonlocal}Same as in Fig.~\ref{Fig:inverselocal} but now for a 40~nm thick TiN film described by the nonlocal KR in-plane EM response function given by Eqs.~(\ref{eps2}) and (\ref{omegapk}). Shown are the $n = 0$ fundamental plasma mode (lower dashed curve) and the $n = 1$ standing wave (upper dashed curve) of Eq.~(\ref{Eq:Standingwavesnonlocal}) as well as the Brewster mode of Eq.~(\ref{Eq:Brewsternonlocal}) (solid line for p-polarization). The behavior of all modes can be seen to drastically change for small angles of incidence (cf. also Fig.~\ref{fig01}).}
\end{figure}

\begin{figure}
\includegraphics[width=0.49\textwidth]{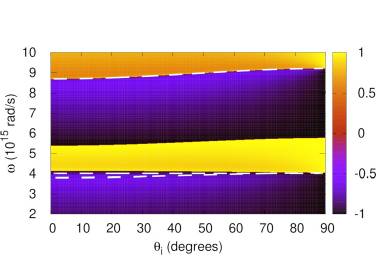}
\includegraphics[width=0.49\textwidth]{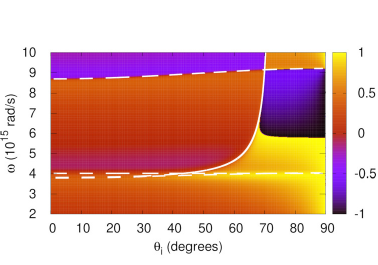}
\caption{\label{Fig:phaselocal}Phase (normalized by $\pi$) for the reflection coefficients whose inverse squares are shown in Fig.~\ref{Fig:inverselocal}. The dashed and solid lines are the same as in Fig.~\ref{Fig:inverselocal}. The phase jumps can be seen at the $n = 1$ standing wave mode (upper dashed line) for both s- and p-polarizations. Extra phase jumps occur between the dashed lines and close to the Brewster mode (solid line) for s- and p-polarizations, respectively.}
\end{figure}

\begin{figure}
\includegraphics[width=0.49\textwidth]{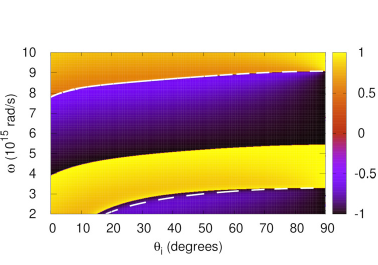}
\includegraphics[width=0.49\textwidth]{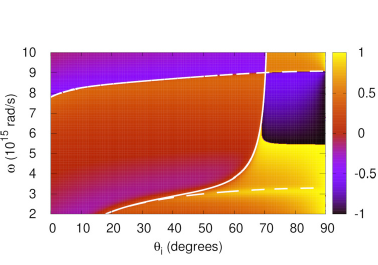}
\caption{\label{Fig:phasenonlocal}Phase (normalized by $\pi$) for the reflection coefficients whose inverse squares are shown in Fig.~\ref{Fig:inversenonlocal}. The dashed and solid lines are the same as in Fig.~3. The impact of the EM response nonlocality can be seen at small angles of incidence.}
\end{figure}

\begin{figure}
\includegraphics[width=0.49\textwidth]{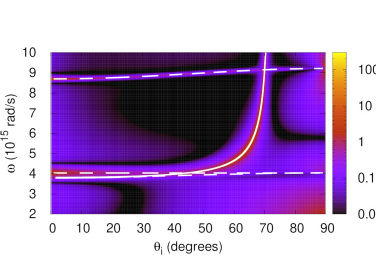}
\includegraphics[width=0.49\textwidth]{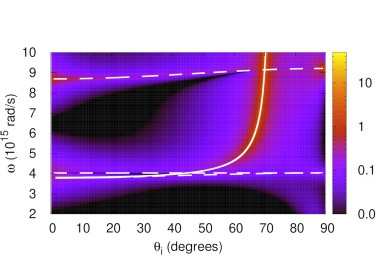}
\caption{\label{Fig:GHlocal}GH shifts $\Delta_{\rm GH}$ in $\mu$m (left) and $\Theta_{\rm GH}$ in mrad (right) for p-polarized light impinging on a 40 nm thick TiN film with inverse reflectivity and phase shown in Figs.~\ref{Fig:inverselocal} and \ref{Fig:phaselocal}, respectively. The solid and dashed lines are the same as in the right panel of Fig.~\ref{Fig:inverselocal}. It can be seen that the lateral shift $\Delta_{\rm GH}$ (left) is particularly large for the Brewster mode (solid line) and the standing wave solutions with $n = 0$ (fundamental plasma mode) and $n = 1$, as well as for the Christiansen point (at small angels of incidence). The angular shift $\Theta_{\rm GH}$ (right) is particularly large when the Brewster mode condition (solid line) and the $n = 1$ standing wave condition (upper dashed line) are fulfilled.}
\end{figure}

\begin{figure}
\includegraphics[width=0.49\textwidth]{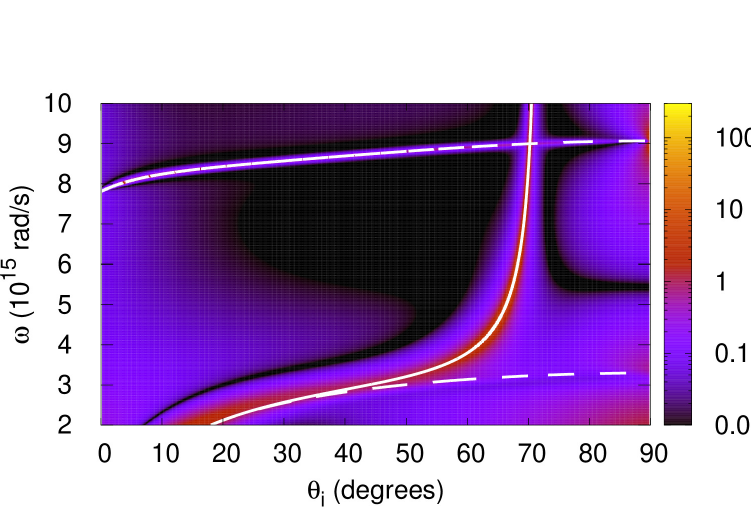}
\includegraphics[width=0.49\textwidth]{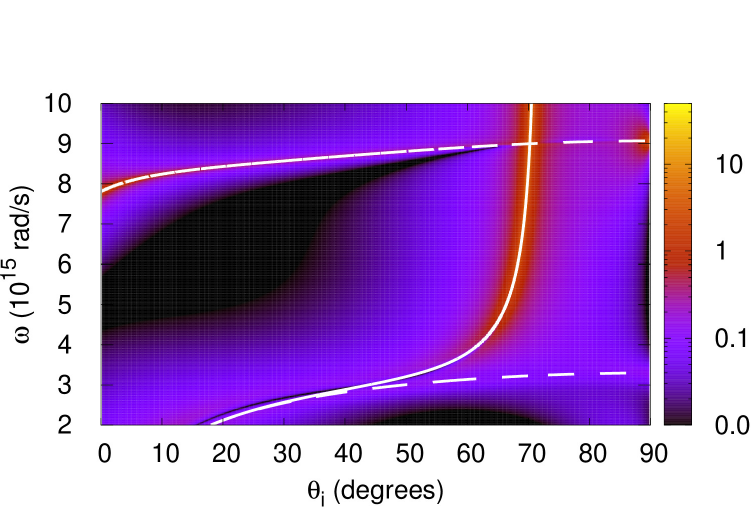}
\caption{\label{Fig:GHnonlocal}Same as in Fig.~\ref{Fig:GHlocal} for p-polarized light impinging on a 40 nm thick TiN film with inverse reflectivity and phase shown in Figs.~\ref{Fig:inversenonlocal} and \ref{Fig:phasenonlocal}, respectively. The lines are the same as in Fig.~\ref{Fig:inversenonlocal}. The impact of the EM response nonlocality can be seen at small angles of incidence.}
\end{figure}

It should be noted that Eq.~(\ref{Eq:Standingwavesnonlocal}) can be generalized to include dissipative effects, too, if one starts from the constraint $|R_p|^2=0$ as given by Eq.~(\ref{Rp}) with $r_p^{12}=-r_p^{23}$ canceled out to exclude the interface modes already discussed. Then,
\begin{equation}
1+\re^{2 \ri (\gamma_2-\gamma_2^\ast) d}=2\mbox{Re}\big(\re^{-2\ri \gamma_2^\ast d}\big),
\label{standingwaves}
\end{equation}
which can be brought to the trigonometrical form
\[
\sin\!\big[(\gamma_2^\prime+\gamma_2^{\prime\prime})d\big]\sin\!\big[(\gamma_2^\prime-\gamma_2^{\prime\prime})d\big]=0
\]
yielding
\begin{equation}
(\gamma_2^\prime\pm\gamma_2^{\prime\prime})d=\pi n,\;\;\;n=0,\pm1,\pm2,\pm3,...,
\label{standwaves}
\end{equation}
where $\gamma_2^\prime$ and $\gamma_2^{\prime\prime}$, the real and imaginary parts of $\gamma_2$, can be obtained from its complex exponential form
\begin{eqnarray}
\gamma_2 = k_0\sqrt{\epsilon_{\rm TiN} - (k/k_0)^2}=\gamma_2^\prime+\ri\gamma_2^{\prime\prime}=|\gamma_2|\,\re^{\ri({\rm Arg}(\gamma_2)+2\pi m)/2},\;m=0,1\hskip0.5cm\nonumber\\
|\gamma_2|=k_0\Big\{\big[\epsilon_{\rm TiN}^\prime\!-\!(k/k_0)^2\big]^2+\epsilon_{\rm TiN}^{\prime\prime\,2}\Big\}^{\!1/4}\!\!\!,\;\;\;\;\;
{\rm Arg}(\gamma_2)=\arctan\!\left[\frac{\epsilon_{\rm TiN}^{\prime\prime}}{\epsilon_{\rm TiN}^\prime\!-\!(k/k_0)^2}\right].\nonumber
\end{eqnarray}
Plugging them in Eq.~(\ref{standwaves}) leads after straightforward simplifications to the transcendental equation as follows
\begin{equation}
\sin\!\left\{\frac{1}{2}\arctan\!\left[\frac{\epsilon_{\rm TiN}^{\prime\prime}}{\epsilon_{\rm TiN}^\prime\!-\!(k/k_0)^2}\right]\pm\frac{\pi}{4}\right\}=
\frac{\pi n}{\sqrt{2}k_0d\,\Big\{\big[\epsilon_{\rm TiN}^\prime\!-\!(k/k_0)^2\big]^2+\epsilon_{\rm TiN}^{\prime\prime\,2}\Big\}^{\!1/4}},\;n=0,\pm1,\pm2,\pm3,...
\label{standing}
\end{equation}

The transcendental equation (\ref{standing}) sets up the EM modes of the TD film that are responsible for its zero reflection and thus for either enhanced transmission or enhanced absorption of external EM radiation incident on the film. It can be solved for $\omega$ analytically. In the negligible dissipation case, one has $\epsilon_{\rm TiN}^{\prime\prime}\!\ll\!|\epsilon_{\rm TiN}^\prime\!-\!(k/k_0)^2|$ to obtain
\[
k_0\sqrt{|\epsilon_{\rm TiN}^\prime\!-\!(k/k_0)^2|}=\frac{\pi n}{d},\;\;\;n \in \mathds{N},
\]
which after plugging $\epsilon_{\rm TiN}^\prime(\omega,k)$ in it leads to the generalized form of Eq.~(\ref{Eq:Standingwavesnonlocal}) as follows
\begin{equation}
\omega = \sqrt{\frac{k^2 c^2}{\epsilon_b} + \omega_p^2(k) + \mbox{sign}\!\left[\omega^2-\omega_p^2(k)-(kc/\!\sqrt{\epsilon_b})^2\right]
\frac{c^2}{\epsilon_b}\biggl( \frac{\pi n}{d} \biggr)^2},\;\;\;n \in \mathds{N}.
\label{genstanding}
\end{equation}
This includes both propagating waves of Eq.~(\ref{Eq:Standingwavesnonlocal}) and evanescent waves as well, in medium 2 (TiN film), which come out as solution branches with $\omega^2\!>\omega_p^2+(kc/\!\sqrt{\epsilon_b})^2$ and $\omega^2\!<\omega_p^2+(kc/\!\sqrt{\epsilon_b})^2$, respectively. In the strong dissipation case, where $\epsilon_{\rm TiN}^{\prime\prime}\!\gg\!|\epsilon_{\rm TiN}^\prime\!-\!(k/k_0)^2|$, the properties of the arctangent allow one to rewrite Eq.~(\ref{standing}) in the form
\[
\sin\!\left\{\frac{\pi}{4}\,\mbox{sign}\!\left[\epsilon_{\rm TiN}^\prime\!-\!(k/k_0)^2\right]\pm\frac{\pi}{4}\right\}=\frac{\pi n}{k_0d\sqrt{2\epsilon_{\rm TiN}^{\prime\prime}}},
\]
with the only legitimate solution (the solution that stays asymptotically correct as $d\rightarrow\infty$) given by $\epsilon_{\rm TiN}^\prime\!-\!(k/k_0)^2=0$ and $n=0$. With our nonlocal $\epsilon_{\rm TiN}^\prime(\omega,k)$ of Eqs.~(\ref{eps2})-(\ref{epstin}), this yields the dispersion equation for the fundamental plasma mode of the finite-thickness TD film
\[
\omega = \sqrt{\frac{k^2 c^2}{\epsilon_b} + \omega_p^2(k)},\;\;\;n=0,
\]
which can be combined with Eq.~(\ref{genstanding}) to give the final standing wave solution as follows
\begin{equation}
\omega = \sqrt{\frac{k^2 c^2}{\epsilon_b} + \omega_p^2(k) + \mbox{sign}\!\left[\omega^2-\omega_p^2(k)-(kc/\!\sqrt{\epsilon_b})^2\right]
\frac{c^2}{\epsilon_b}\biggl( \frac{\pi n}{d} \biggr)^2},\;\;\;n=0,1,2,3,...
\label{genstandingfin}
\end{equation}
However, one has to remember that here, contrary to the $n\ne0$ low-dissipative modes of Eq.~(\ref{genstanding}), the fundamental mode with $n=0$ is associated with strong dissipation of EM radiation absorbed by the TD film to generate the in-plane plasma waves in the system.

Figure~\ref{fig01} shows and describes in the caption the features of the propagating and evanescent wave dispersion relations given by Eq.~(\ref{genstandingfin}) with $n=0,1,2$ and $3$, presented in the dimensionless $(\omega,k)$-space for a few free standing TiN films of decreasing thickness. The Brewster and surface modes of Eq.~(\ref{Eq:Brewsternonlocal}) are also shown. Note that due to the in-plane reflection symmetry of the free standing TD film system, all modes shown lead to zero $p$-wave reflection coefficient in Eq.~(\ref{Rp}) as it follows from the discussion above. The detailed analysis and general properties of the evanescent wave solutions in ultrathin TD plasmonic films can be found in Refs.~\cite{BondPRR20,BondAP2}. The GH shift calculations require the knowledge of the propagating wave solutions which we are therefore focusing on below.

Figures~\ref{Fig:inverselocal} and \ref{Fig:inversenonlocal} show the calculated inverse reflectivities and discuss in the captions some of the propagating wave dispersion relations in the $(\omega,\theta_i)$-space, described by local (Drude) and nonlocal KR in-plane EM response functions as given by Eqs.~(\ref{eps2}) and (\ref{omegapk}) for infinitely large and finite $d$, respectively. It can be seen that the dispersion relations of Eqs.~(\ref{Eq:Brewsternonlocal}) and (\ref{Eq:Standingwavesnonlocal}) for the Brewster modes and standing wave modes with losses neglected are in full agreement with direct numerical calculations of Eq.~(\ref{Rp}) including losses. It can also be seen that at small angles of incidence, i.e.\ for small $k$, the nonlocality of the in-plane EM response plays an important role. More features in the $(\omega,\theta_i)$-space can be seen in Figs.~\ref{Fig:phaselocal}--\ref{Fig:GHnonlocal}. Figures~\ref{Fig:phaselocal} and \ref{Fig:phasenonlocal} show the phases (normalized by $\pi$) of the reflection coefficients whose inverse squares are shown in Figs.~\ref{Fig:inverselocal} and \ref{Fig:inversenonlocal}, respectively. Figures~\ref{Fig:GHlocal} and \ref{Fig:GHnonlocal} present the respective GH shifts, where it can be seen that due to the EM response nonlocality the large GH shifts can be obtained for frequencies below the bulk plasma frequency $\omega_p^{3D}$, for example, by using He-Ne laser light.

\begin{figure}
\includegraphics[width=0.75\textwidth]{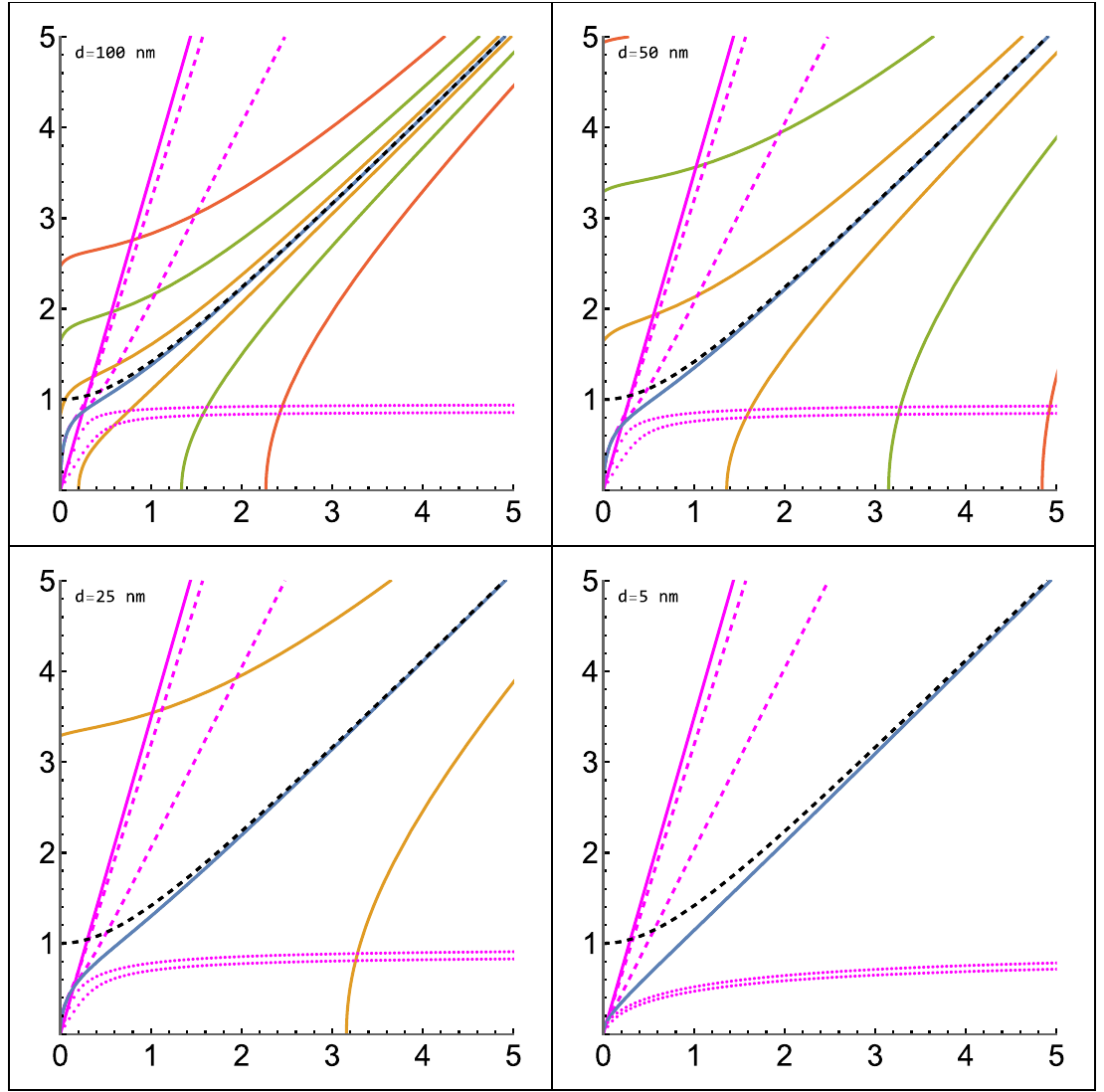}
\caption{Same as in Fig.~\ref{fig01} for air/TiN/MgO TD films. Purple dashed and dotted lines show the split-up non-degenerate Brewster and surface modes, respectively, given by Eq.~(\ref{Eq:Brewsternonlocal}) for the air/TiN interface and by Eq.~(\ref{Eq:Brewsternonlocal23}) for the TiN/MgO interface. Purple solid line is the Brewster mode of the hypothetical air/MgO interface introduced in 4.), discussed in A)--C), and referred to as zero Brewster mode (zBM) in the main text. The zBM intersection points with standing wave modes (including the $n=0$ fundamental plasma mode) are the phase singularity points (or points of topological darkness) to yield $R_p=0$ and thus to greatly enhance the GH effect.}
\label{fig02}
\end{figure}

\begin{figure}
\includegraphics[width=0.8\textwidth]{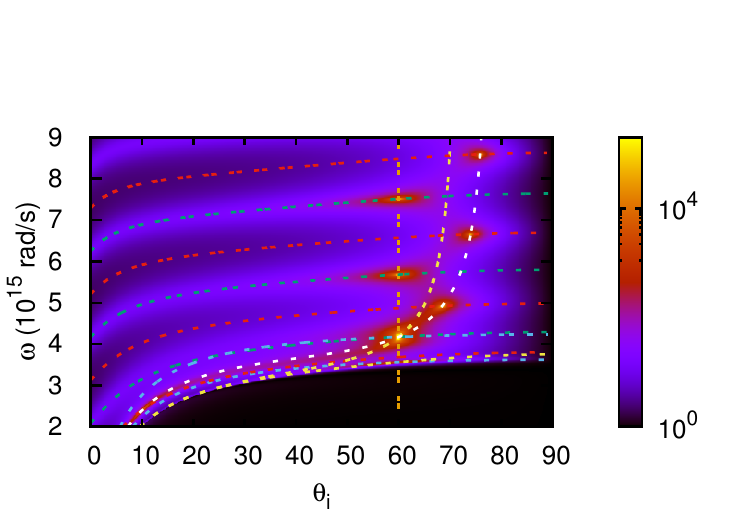}
\caption{\label{Fig:rpplasma}Inverse reflectivity $1/|R_p|^2$ for a 150 nm thick TiN film on MgO calculated using the nonlocal KR in-plane EM response as given by Eqs.~(\ref{eps2}) and (\ref{omegapk}) with losses neglected by setting $\Gamma_D = 0$. The figure presents the features listed in the text. They are: 1.)~The air/TiN interface Brewster mode is shown by the (upper) yellow dashed line; 2.)~the MgO/TiN interface Brewster mode is shown by the (lower) yellow dashed line; 3.)~the standing wave solutions for $n = 1,2,3$ from Eq.~(\ref{Eq:Standingwavesnonlocal}) are shown by the green dashed lines; 4.)~the vertical dashed orange line indicates the hypothetical MgO/Air interface Brewster mode (also called zBM in the main text); 5.)~and 6.)~the blue dashed lines mark the two Christiansen points. Additionally presented are the generalized Brewster mode (gBM) from F) by the white dashed line and the standing wave solutions with $n = 0.5, 1.5, 2.5, 3.5$ from Eq.~(\ref{Eq:Standingwavesnonlocal}), or the condition in F), by the red dashed lines. The reflection zeros can be seen at the crossing points of the standing waves from 3.) (green dashed lines) and the zBM from 4.) (vertical dashed line), as described in B) in the text. More reflection zeroes can be seen at the intersection of the standing wave solutions from F) (red dashed lines) and the gBM from F). All the Brewster modes can be seen to intersect with one of the Christiansen points and the $n = 1$ standing wave mode. There is also the intersection of the Brewster modes 2.), 3.) with the Christiansen point at the 60$^\circ$ incidence angle and $\omega = 3.55\times10^{15}\,{\rm rad/s}$, to provide another reflection zero (barely seen) corresponding to A) and D) listed in the text. All these intersection points provide the phase singularities to greatly enhance the GH shifts.}
\end{figure}

\begin{figure}
\includegraphics[width=0.8\textwidth]{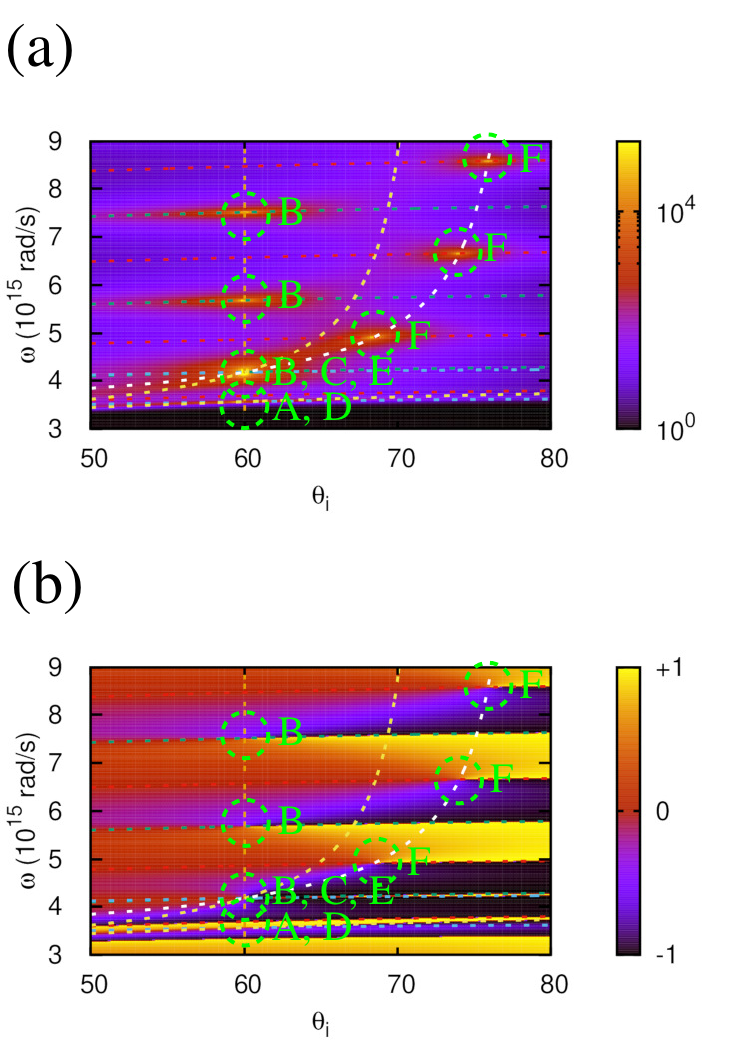}
\caption{\label{Fig:phasesingularities}(a)~Inverse reflectivity $1/|R_p|^2$ as in Fig.~\ref{Fig:rpplasma} but now for a 40 nm thick TiN film on MgO and (b)~the phase $\pi$-normalized of the respective reflection coefficient. The lines are the same as in Fig.~\ref{Fig:rpplasma}. The labeled green circles mark the mode intersection points and the phase singularities as per the classification A)--F) in the text, to enhance the respective GH shifts.}
\end{figure}

\subsection{Broken In-Plane Reflection Symmetry: TiN Film on MgO Substrate}

In this case, there are two different (inequivalent) interfaces with their respective interface modes. Our TD plasmonic TiN film (medium 2) is now sandwiched between air (medium 1) and a MgO substrate with $\epsilon_{\rm MgO} = 3.0$ (medium 3). The top-bottom interface mode degeneracy is lifted as compared to the free standing plasmonic film case. However, we show in what follows that the propagating standing waves of the film can still provide zero reflection given that proper Brewster mode constraints are fulfilled at both of the inequivalent interfaces, which is possible at mode intersection points in the $(\omega,k)$-space. They are the (topological) phase singularity points to replace the lines of the free standing film case and thus to greatly enhance the GH effect.

For the air/TiN interface Brewster and surface modes one still has Eq.~(\ref{Eq:Brewsternonlocal}), whereas for the MgO/TiN interface
\begin{equation}
	\omega^2 = \frac{1}{2}\biggl(k^2c^2 \frac{\epsilon_b + \epsilon_3}{\epsilon_b \epsilon_3} + \omega_p^2(k) \biggr) \pm \sqrt{\frac{1}{4}\biggl(k^2 c^2 \frac{\epsilon_b + \epsilon_3}{\epsilon_b \epsilon_3} + \omega_p^2(k) \biggr)^2 - \frac{k^2 c^2 \omega_p^2(k)}{\epsilon_3}},
	\label{Eq:Brewsternonlocal23}
\end{equation}
where $+(-)$ correspond to the Brewster (surface) mode in the propagating (evanescent) wave region of medium 3 and $\epsilon_3=\epsilon_{\rm MgO}$. The propagating modes of the TiN film itself are still given by the standing wave solutions of Eq.~(\ref{Eq:Standingwavesnonlocal}). However, contrary to thick films where $\omega_p(k)\!\sim\!\omega_p^{3D}$ is the same for all modes, for the TD plasmonic film system $\omega_p(k)\!\sim\!\omega_p^{3D}\sqrt{kd\epsilon_b/(\epsilon_1+\epsilon_3)}$ as per Eq.~(\ref{omegapk}), which is different from $\omega_p(k)$ of the free standing TD film case.\\

All modes of relevance to the GH effect in our system can be summarized as follows:
\begin{enumerate}
\item[1.)] The air/TiN interface Brewster mode described by the $+$ sign branch of Eq.~(\ref{Eq:Brewsternonlocal}), to yield $r^{12}_{\rm p}=0$;

\item[2.)] The MgO/TiN interface Brewster mode described by the $+$ sign branch of Eq.~(\ref{Eq:Brewsternonlocal23}), to yield $r^{23}_{\rm p}=0$;

\item[3.)] The propagating modes of the TiN film given by Eq.~(\ref{genstandingfin}), originating from Eq.~(\ref{standingwaves}) with $\gamma_2^{\prime\prime}=0$, to yield $1 - \re^{2 \ri \gamma_2 d} = 0$ for $\gamma_2=\pi n/d$ with $n=0,1,2,...$;

\item[4.)] For ultrathin TD films, of significance can also be the Brewster mode of a hypothetical interface between media 1 and 3, in which case by analogy with Eq.~(\ref{Eq:Brewster}) one has $k=(\omega/c)\sqrt{\epsilon_3/(\epsilon_{3} + 1)}$, or $\omega=kc\sqrt{(\epsilon_3 + 1)/\epsilon_3}$, to yield $r^{13}_{\rm p} = 0$;

\item[5.)] From Eq.~(\ref{Rp}) it can be seen that $\epsilon_2' (k) = 1$ (air/TiN interface Christiansen point), or $\omega=\omega_p(k)\sqrt{\epsilon_b/(\epsilon_b-1)}$ as per Eq.~(\ref{omegapk}), leads to $r^{12}_{\rm p} = 0$;
		
 \item[6.)] Similarly, $\epsilon_2'(k)=\epsilon_3$ (MgO/TiN interface Christiansen point), or $\omega=\omega_p(k)\sqrt{\epsilon_b/(\epsilon_b-\epsilon_3)}$, leads to $r^{23}_{\rm p} = 0$.
\end{enumerate}

There is a simple method to find the zeroes of the reflection coefficient in Eq.~(\ref{Rp}). As they come from the numerator
\begin{equation}
N(\omega,k) = r_p^{12} + r_p^{23} \re^{2 \ri \gamma_2 d},
\label{Nomegak}
\end{equation}
to find them we start with the Brewster mode of a hypothetical interface between medium 1 and 3 (air/MgO) mentioned in 4.) above. This is the case where $\gamma_1 \epsilon_3 = \gamma_3 \epsilon_1$. Using this inside $r_p^{13}=0$ leads to $r_p^{12} = - r_p^{23}$, now for TD plasmonic film systems with broken in-plane reflection symmetry. This is possible because $r_p^{12}$ and $r_p^{23}$ are functions of $\omega$ and $k$, and their equality implies nothing but their intersection point in the $(\omega,k)$-space as opposed to their exact coincidence (identity) in the degenerate case of the preserved in-plane reflection symmetry of free standing TD films. Hence, one has
\begin{equation}
N(\omega,k) = r_p^{12} + r_p^{23} \re^{2 \ri \gamma_2 d} = r_p^{23} \bigl(-1 + \re^{2 \ri \gamma_2 d}\bigr).
\end{equation}
This expression is now zero in the following cases:
\begin{enumerate}
\item[A)] If we assume that $r_p^{13} = 0$ and $\gamma_1 \epsilon_3 = \gamma_3 \epsilon_1$ are fulfilled then $N(\omega,k) = 0$ if $r_p^{23} = 0$. That means the reflection coefficient $R_p$ can only be zero at the crossing points of the Brewster mode of the air/MgO interface from 4.) and the Brewster mode of the MgO/TiN interface from 2.).
\item[B)] If we assume that $r_p^{13} = 0$ and $\gamma_1 \epsilon_3 = \gamma_3 \epsilon_1$ are fulfilled then $N(\omega,k) = 0$ if $\bigl(-1 + \re^{2 \ri \gamma_2 d}\bigr) = 0$. That means the reflection coefficient $R_p$ can only be zero at the crossing points of the Brewster mode of the air/MgO interface from 4.) and the standing wave modes from 3.).
\item[C)] From the condition A) that when $r_p^{13} = 0$ and $\gamma_1 \epsilon_3 = \gamma_3 \epsilon_1$ are fulfilled and  $r_p^{23} = 0$ then $R_p = 0$ it follows that also when $r_p^{13} = 0$ and $r_p^{12} = 0$ are fulfilled there is zero of $R_p$, because $r^{23}_p = - r_p^{12}$. This means that when the Brewster mode of the  air/MgO interface from 4.) and the Brewster mode of the air/TiN interface from 1.) cross then there is a zero reflection point.

Additionally, one has:

\item[D)] Since  $r_p^{12} = 0$ is also fulfilled for the Christansen point in 5.) there is another possible zero at the crossing point of the Christiansen mode from 5.) and the Brewster mode of the  MgO/TiN interface from 2.).
\item[E)] Since  $r_p^{23} = 0$ is also fulfilled for the Christansen point in 6.) there is another possible zero at the crossing point of the Christiansen mode from 6.) and the Brewster mode of the  MgO/TiN interface from 2.).
\item[F)] The above solutions come from either intersection of the Brewster and Christiansen modes or the intersection of the Brewster mode of the (hypothetical) medium~1/medium~3 interface and the standing waves of medium~2. On closer inspection of Eq.~(\ref{Nomegak}), however, one finds another class of solutions which is associated not with the standing waves fulfilling $\bigl(-1 + \re^{2 \ri \gamma_2 d}\bigr) = 0$ but with those fulfilling $\bigl(1 + \re^{2 \ri \gamma_2 d}\bigr) = 0$. This comes from the constraint $r_p^{12} = r_p^{23}$ which can be the case when $\gamma_2^2 \epsilon_1 \epsilon_3 = \epsilon_2^2 \gamma_3 \gamma_1$. It leads to the analogue of Eq.~(\ref{Eq:Standingwavesnonlocal}) with $n = 0.5, 1.5, 2.5, \ldots$
\end{enumerate}

Figure~\ref{fig02} shows and comments in the caption on how cases A), B) and C) can be understood in terms of the mode intersection points in the $(\omega,k)$-space. Figures~\ref{Fig:rpplasma} and \ref{Fig:phasesingularities} show and comment on the above listed singularity locations in the inverse reflectivity and reflection coefficient phase in the $(\omega,\theta_i)$-space.

Note that cases B) and F) above are referred to as cases 1 and 2 in the main text. In these two cases the phase singularities are defined by the intersection of the standing wave mode dispersion curves with the two different Brewster mode dispersion curves. Note also that in our configuration cases C) and E) coincide with the energetically lowest phase singularity of case 1 in the main text, or B) here, which is red-shifted for thinner films due to the nonlocal in-plane EM response effect. Moreover, cases A) and D) herein define another phase singularity not covered by cases B) and F), or cases 1 and 2 in the main text.

Finally, we would like to stress that the Brewster modes discussed here to provide the most important singularity points are those given by the reflection coefficient zeros (also known as improper modes~\cite{Oliner93}), to which therefore it is impossible to assign a group velocity. In contrast, the surface modes are those given by the poles of the reflection coefficient~\cite{BondPRR20}. They are the proper eigen modes confined to the interface~\cite{BondAP2}, to which one can assign both phase and group velocity. The differences between the two can be clearly seen in Fig.~\ref{fig01} and Fig.~\ref{fig02} for the TD films with and with no in-plane reflection symmetry, respectively.
